\documentclass{amsart}
\usepackage{amsmath}
\theoremstyle{plain}
\newtheorem{theorem}{Theorem}
\newtheorem{proposition}[theorem]{Proposition}
\newtheorem{corollary}[theorem]{Corollary}
\newtheorem{lemma}[theorem]{Lemma}

\theoremstyle{remark}
\newtheorem*{remark*}{Remark}

\theoremstyle{definition}

\numberwithin{equation}{section}

\newcommand\CC{{\mathbf C}}
\newcommand\into{\int_\Omega}
\newcommand\OmegaN{{\Omega_{\fam0 norm}}}
\newcommand\intoN{\int_\OmegaN}
\newcommand\RR{{\mathbf R}}
\newcommand\spr[1]{\langle#1\rangle}
\newcommand\oz{{\overline z}}
\newcommand\RE{\operatorname{Re}}
\newcommand\dbar{\overline\partial}
\newcommand\HH{\mathcal H}
\newcommand\oy{{\overline y}}
\newcommand\wt[1]{\widetilde{#1}}
\newcommand\wtt[1]{\wt{\wt{#1}}}
\newcommand\bP{\mathbf P}
\newcommand\Th{T^{(h)}}
\newcommand\CCN{{\CC^N}}
\newcommand\CCNN{{\CC^{N\times N}}}
\newcommand\Tr{\operatorname{Tr}}
\newcommand\diag{\operatorname{diag}}
\newcommand\bbr[1]{\{\!\{#1\}\!\}}
\newcommand\LL{\mathcal L}
\newcommand\od{\overline d}
\newcommand\ou{\overline u}
\newcommand\oc{\overline c}
\newcommand\UN{{U(N)}}
\newcommand\sumr{\sum_{r=0}^\infty}
\newcommand\hol{{\text{\rm hol}}}
\newcommand\bbk[1]{{\mathbf[\!\mathbf[#1\mathbf]\!\mathbf]}}
\newcommand\MM{\mathcal M}
\newcommand\ow{\overline w}
\newcommand\bep{\mathbf\epsilon}
\def\Sb#1\endSb{_{\begin{subarray}{c}#1\end{subarray}}}

\newcommand\DD{\mathbf D}
\allowdisplaybreaks[4]

\begin{document}

\title[Berezin-Toeplitz quantization]
{Berezin-Toeplitz quantization over matrix domains}
\author{S. Twareque Ali, M.~Engli\v s}
\address{Mathematics Institute, \v Zitn\'a 25, 11567 Prague 1, Czech Republic}
\email{englis{@}math.cas.cz}
\thanks{The first author would like to acknowledge support from the NSERC,
Canada and FQRNT, Qu\'ebec. Research of the second author was supported by
GA~\v CR grant no.~201/03/0041 and AV~\v CR research plan no.~AV0Z10190503.}
\address{Department of Mathematics and Statistics, Concordia University,
Montr\'eal, Qu\'ebec, Canada H4B 1R6}
\email{stali{@}mathstat.concordia.ca}
\begin{abstract} We~explore the possibility of extending the well-known
Berezin-Toeplitz quantization to reproducing kernel spaces of vector-valued
functions. In~physical terms, this can be interpreted as accommodating the
internal degrees of freedom of the quantized system. We~analyze in particular
the vector-valued analogues of the classical Segal-Bargmann space on the
domain of all complex matrices and of all normal matrices, respectively,
showing that for the former a semi-classical limit, in the traditional sense,
does not exist, while for the latter only a certain subset of the quantized
observables have a classical limit: in~other words, in~the semiclassical limit
the internal degrees of freedom disappear, as~they should. We~expect that a
similar situation prevails in much more general setups. \end{abstract}

\maketitle

\section{Introduction}\label{INTR}
Let $\Omega$ be a symplectic manifold, with symplectic form $\omega$, and $\HH$
a subspace of $L^2(\Omega,d\mu)$, for~some measure~$\mu$, admitting a
reproducing kernel~$K$. For $\phi\in C^\infty(\Omega)$, the \emph{Toeplitz
operator} $T_\phi$ with symbol $\phi$ is the operator on $\HH$ defined~by
$$ T_\phi f = \bP(\phi f),  \qquad f\in\HH,  $$
where $\bP:L^2(\Omega,d\mu)\to\HH$ is the orthogonal projection. Using the
reproducing kernel~$K$, this can also be written~as
$$ T_\phi f(x) = \into f(y)\phi(y) K(x,y)\,d\mu(y).  $$
It~is easily seen that $T_\phi$ is a bounded operator whenever $\phi$ is a
bounded function, and $\|T_\phi\|_{\HH\to\HH}\le\|\phi\|_\infty$, the supremum
norm of~$\phi$.

Suppose now that both the measure $\mu$ and the reproducing kernel subspace
$\HH$ are made to depend on an additional parameter~$h>0$ (shortly to be
interpreted as the Planck constant), in~such a way that the associated Toeplitz
operators $\Th_\phi$ on $\HH_h$ satisfy, as~$h\searrow0$,
\begin{equation}  \|\Th_\phi\|_{\HH_h\to\HH_h} \to \|\phi\|_\infty,
\label{tag:TNUL} \end{equation}
and
\begin{align}
& \|\Th_\phi \Th_\psi - \Th_{\phi\psi} \|_{\HH_h\to\HH_h} \to 0,
\label{tag:TA}  \\
& \|\tfrac{2\pi}{ih}[\Th_\phi,\Th_\psi]- \Th_{\{\phi,\psi\}} \|
 _{\HH_h\to\HH_h} \to 0  \label{tag:TB}  \end{align}
(where $\{\cdot,\cdot\}$ is the Poisson bracket with respect to~$\omega$), and,
more generally,
$$ \Th_\phi \Th_\psi \approx \sum_{j=0}^\infty h^j \; \Th_{C_j(\phi,\psi)}
\qquad\text{as } h\to0,  $$
for some bilinear differential operators $C_j:C^\infty(\Omega)\times C^\infty
(\Omega)\to C^\infty(\Omega)$, with $C_0(\phi,\psi)=\phi\psi$ and $C_1(\phi,
\psi)-C_1(\psi,\phi)=\frac i{2\pi}\{\phi,\psi\}$. Here the last asymptotic
expansion means, more precisely, that
\begin{equation}  \Big\| \Th_\phi \Th_\psi - \sum_{j=0}^N h^j
\Th_{C_j(\phi,\psi)} \Big\|
= O(h^{N+1}) \quad\text{as } h\searrow0 , \qquad \forall N=0,1,2,\dots.
\label{tag:TC}  \end{equation}
One then speaks of the \emph{Berezin-Toeplitz quantization.} Indeed, it~is well
known that the recipe
$$ \phi * \psi := \sum_{j=0}^\infty h^j\, C_j(\phi,\psi)  $$
then gives a \emph{star-product} on ~$\Omega$, and (\ref{tag:TA}),
(\ref{tag:TB}) just amount to its correct semiclassical limit.

Observe that if we introduce the normalized reproducing kernels ---
or~\emph{coherent states} ---~by
$$ k_y := \frac{K(\cdot,y)}{\|K(\cdot,y)\|_{\HH}}, \qquad\text{i.e.}\qquad
k_y(x) = \frac{K(x,y)}{K(y,y)^{1/2}},  $$
and define the \emph{Berezin transform} of an operator $T$ on~$\HH$~by
$$ \wt T(y) := \spr{T k_y,k_y}_\HH,  $$
then (\ref{tag:TC}) implies that as $h\to0$,
\begin{equation}  \wt{\Th_\phi \Th_\psi} \approx \sum_{j=0}^\infty h^j \;
\wt{\Th_{C_j(\phi,\psi)}},  \label{tag:TE}  \end{equation}
pointwise and even uniformly on~$\Omega$.
Often, one also has a stronger version of (\ref{tag:TNUL}), namely
\begin{equation}  \wt{\Th_\phi}(x) \to \phi(x) \qquad\forall x\in\Omega.
\label{tag:TF}  \end{equation}

The~simplest instance of the above situation is $\Omega=\RR^{2n}\simeq\CC^n$,
with the standard (Euclidean) symplectic structure, and
\begin{equation}  \HH_h = L^2_\hol(\Omega,d\mu_h)   \label{tag:SB}
\end{equation}
the Segal-Bargmann space of all holomorphic functions square-integrable with
respect to the Gaussian measure $d\mu_h(z):=e^{-\|z\|^2/h} (\pi h)^{-n}\,dz$
($dz$~being the Lebesgue measure on~$\CC^n$). The space $\HH_h$ admits the
reproducing kernel $K_h(x,y)=e^{\spr{x,y}/h}$. The Berezin transform turns out
to be given just by the familiar heat operator semigroup,
$$ \wt{\Th_\phi} (x) = (\pi h)^{-n} \int_{\CC^n} e^{-\|x-y\|^2/h}
\phi(y) \,dy ,  $$
so~that, by~the stationary phase (WJKB) expansion
(see Section~\ref{STPH} below; for a discussion of the (WJKB) approximation,
see for example \cite{Me}, Chapter~7)
\footnote{Throughout this paper, we~are using the slightly nonstandard
Laplacian $\Delta=\sum_j \partial^2/\partial z_j \partial\oz_j$, which differs
from the usual one by a factor of~4.}
\begin{equation}  \wt{\Th_\phi}(x) = \sum_{j=0}^\infty \frac{h^j}{j!} \Delta^j
\phi(x) . \label{tag:TD}  \end{equation}
Similarly,
$$ \wt{\Th_\phi \Th_\psi}(x) = (\pi h)^{-2n} \int_{\CC^n} \int_{\CC^n}
\phi(y)\psi(z) e^{(\spr{x,y}+\spr{y,z}+\spr{z,x}-\|x\|^2-\|y\|^2-\|z\|^2)/h}
\,dy\,dz ,  $$
so~by stationary phase again,
\begin{equation}  \wt{\Th_\phi \Th_\psi}(x) = \sum_{\alpha,\beta,\gamma}
\frac{\partial^\alpha \dbar^{\alpha+\gamma}\phi(x)}{\alpha!\gamma!} \;
\frac{\partial^{\beta+\gamma} \dbar^\beta \psi(x)}{\beta!} \;
h^{|\alpha|+|\beta|+|\gamma|}.   \label{tag:TY}  \end{equation}
(Here the summation extends over all multiindices $\alpha,\beta,\gamma$,
i.e.~$n$-tuples of nonnegative integers $\alpha=(\alpha_1,\dots,\alpha_n)$,
etc., and~we are using the usual multiindex notations $|\alpha|=\alpha_1+
\dots+\alpha_n$, $\alpha!=\alpha_1!\dots\alpha_n!$, $\partial^\alpha=
\partial^{|\alpha|}/\partial x_1^{\alpha_1}\dots\partial x_n^{\alpha_n}$.)
Inserting (\ref{tag:TD}) and (\ref{tag:TY}) into~(\ref{tag:TE}), we~get
formulas for the cochains~$C_j$:
\begin{equation}  C_j(\phi,\psi) = \sum_{|\alpha|=j} \frac1{\alpha!}
\,\partial^\alpha\phi \,\cdot\, \dbar^\alpha\psi.   \label{tag:CF}
\end{equation}
The resulting star-product coincides, essentially, with the
familiar Moyal product.

Other examples of Berezin-Toeplitz quantization include the unit disc $\DD$
with the Poincar\'e metric, bounded symmetric domains, strictly pseudoconvex
domains with metrics having reasonable boundary behaviour, or,~provided one
allows not only holomorphic functions but also sections of line bundles as
elements of~$\HH_h$, all compact K\"ahler manifolds whose K\"ahler form is
integral. In~all these cases, the choice of the spaces~(\ref{tag:SB}) which
works are the weighted Bergman spaces $\HH_h = L^2_\hol(\Omega,e^{-\Phi/h}
\omega^n)$, where $n$ is the complex dimension of~$\Omega$ and $\Phi$ is a
K\"ahler potential for~$\omega$ (so,~for instance, for~the unit disc $\HH_h
=L^2_\hol(\DD,(1-|z|^2)^{(1/h)-2})$). See~\cite{KS}, \cite{BMS} or~\cite{AE}
for the details and further discussion.

In~this note, we~explore the possibility of extending the above formalism to
reproducing kernel spaces of \emph{vector-valued} functions. In~physical terms,
this can be interpreted as accommodating the \emph{internal degrees of freedom}
of the quantized system.

In~more concrete terms, this means that we again consider, for a given $h>0$,
a~suitable measure $\mu_h$ on~$\Omega$, and a reproducing kernel subspace
$\HH_h\subset L^2_\CCN(\Omega,d\mu_h)$, where the subscript $\CCN$ indicates
that we are now dealing with vector-valued functions taking values in~$\CCN$
for some $N\ge1$.
The reproducing kernel $K_h$ of $\HH_h$ will thus now be a matrix-valued
object, $K_h:\Omega\times\Omega\to\CCNN$. We~can again consider, for~each
$\phi\in C^\infty(\Omega)$, the associated Toeplitz operators, and investigate
the existence of the asymptotic expansion~(\ref{tag:TC}). In~fact, we~can now
even allow \emph{matrix-valued} symbols $\phi\in C^\infty_{\CCNN}(\Omega)$.
By~analogy with the scalar-valued case, one~may again expect appropriate
asymptotic expansions
\begin{align}
& \wt{\Th_\phi} (x) = \sum_{j=0}^\infty L_j\phi(x) \; h^j ,, \label{tag:AL}  \\
& \wt{\Th_\phi \Th_\psi} (x) = \sum_{j=0}^\infty M_j(\phi,\psi)(x)\; h^j,
 \label{tag:AM}  \end{align}
with some differential and bidifferential operators $L_j$ and $M_j$,
respectively, whose comparison would yield~(\ref{tag:TE}), thus suggesting that
(\ref{tag:TC}) is also likely to hold.

Unfortunately, at~this level of generality the results are negative:
we~show that for certain spaces $\HH_h$ as above, which are quite natural
generalizations of the Euclidean situation~(\ref{tag:SB}) to vector-valued
functions, the semiclassical expansions (\ref{tag:AL}), (\ref{tag:AM}) fail
to hold. More specifically,
it~seems that there are asymptotic expansions for $\wt{\Th_\phi}(x)$ and
$\wt{\Th_\phi\Th_\psi}(x)$ in powers of~$h$, but their coefficients do not
depend only on the jets of $\phi$ and $\psi$ at~$x$, but also at other points
(i.e.~are not local operators); besides, in~addition to integer powers of~$h$,
half-integer powers seem to also enter the picture. Finally, (\ref{tag:TNUL})
and (\ref{tag:TF}) may also break down.

However, it~turns out that upon restricting to appropriate domains $\Omega$ and
appropriate classes of
functions $\phi,\psi$, the~situation can be saved completely: namely, the
following picture emerges. The admissible functions $\phi$ can be identified
with functions $f(d_1;d_2,\dots,d_N)$ on $\CC\times\CC^{N-1}$ that are
symmetric in the $N-1$ variables $d_2,\dots,d_N$. Clearly, one can associate
a function $u^\#$ of this form to any function $u:\CC\to\CC$ by the recipe
$u^\#(d_1;d_2, \dots,d_N):=u(d_1)$. We~show that for any functions $f,g$ of
the above form, there exist uniquely determined functions $u_r$ on~$\CC$,
$r=0,1,2,\dots$, such~that
\begin{equation}  \wt{\Th_f \Th_g} \approx \sumr h^r \, \wt{\Th_{u^\#_r}}.
\label{tag:TT}  \end{equation}
Further, the $u_r$~are given by differential expressions involving $f$ and~$g$;
and, finally, if $f$ and $g$ themselves are of the form $v^\#$ and $w^\#$,
respectively, for some functions $v,w$ on~$\CC$, then in fact $u_r=C_r(v,w)$
with the bidifferential operators $C_r$ given by (\ref{tag:CF}) for $n=1$,
i.e.~$\Omega=\CC$.
This suggests the following interpretation: our quantum system has $N$ internal
degrees of freedom, which have no classical couterparts, so that only a subset
of the quantized observables have a classical limit. In~the semiclassical limit
the internal degrees of freedom disappear, as~they should. We~conjecture that
similar quantizations can be carried out by our method in much more general
setups.

The~paper is organized as follows. In~Section~\ref{SPCS}, we~introduce the
spaces $\HH_h$ of vector-valued functions on~$\CCNN$, as~well as their
analogues for the subset of all normal matrices in~$\CCNN$. These spaces have
previously appeared in~\cite{AEG}. In~Section~\ref{BTFM}, we~define a
generalization of the Berezin transform --- which will now also be a
matrix-valued object --- and establish its basic properties.
Sections~\ref{BBFM} and~\ref{BBNM} discuss the semiclassical asymptotic
expansions (\ref{tag:AL})--(\ref{tag:AM}) in the above two settings
of the domains of all matrices and all normal matrices, respectively.
In~Section~\ref{STPH} we present our first result in the positive direction,
by~establishing an asymptotic expansion --- which is, however, of~a highly
non-local nature --- for $\wt{\Th_\phi}$ and $\wt{\Th_\phi \Th_\psi}$ in the
case of the normal matrices. Finally, in~Sections~\ref{SUIF} and~\ref{QUIF} we
introduce our restricted class of observables $\phi$ and establish the
asymptotic expansion~(\ref{tag:TT}).

\medskip

\section{The domains and the spaces}\label{SPCS}
Our first domain is $\Omega=\CCNN$, with the measures
$$ d\mu_h(Z) = e^{-\Tr(Z^*Z)/h} \, (\pi h)^{-N^2} \, dZ ,  $$
where $dZ$ denotes the Lebesgue measure on~$\CCNN$. The measures $\mu_h$ are
normalized to be of total mass~one. The functions $\Psi_j(Z):=Z^j$ satisfy
\begin{equation}  \into Z^{*j} Z^k \,d\mu_h(Z) = \delta_{jk} h^k c_k I
\label{tag:ZZ}  \end{equation}
for some numbers $c_k>0$; see~\cite{AEG}. Explicitly, $c_k$~are given~by
(\cite{Gin}, formula (1.40), and~\cite{Kri})
\footnote{The~authors are grateful to M.~Bertola for this result.}
$$ c_k= \begin{cases}
\dfrac{(k+N+1)!}{N!(k+1)(k+2)} \qquad\text{for } k\ge N-1, \\
\dfrac{(k+N+1)!}{N!(k+1)(k+2)} - \dfrac{N!}{(k+1)(k+2)(N-k-2)!}
\qquad\text{for } k< N-1,  \end{cases}   $$
that is,
\begin{equation}  c_k = \frac1{(k+1)(k+2)} \bigg[ \prod_{j=1}^{k+1} (N+j) -
\prod_{j=1}^{k+1} (N-j) \bigg].   \label{tag:CK}  \end{equation}
In~particular, $c_0=1$, $c_1=N$, $c_2=N^2+1$,~etc.

It~follows that if $\chi_1,\dots,\chi_N$ is the standard basis of~$\CCN$, then
the functions
\begin{equation}  \frac {Z^k \chi_j}{\sqrt{c_k h^k}}, \qquad j=1,\dots,N,
\ k=0,1,2,\dots, \label{tag:ZK}  \end{equation}
are orthonormal in~$L^2_\CCN(\Omega,d\mu_h)$.
Let $\HH_h$ be the subspace spanned by these functions. Then the function
\begin{equation}  K_h(X,Y)= \sum_{k=0}^\infty \frac{X^k Y^{*k}}{c_k h^k}
\label{tag:KH}  \end{equation}
converges for all $X,Y\in\Omega$ and is the reproducing kernel of~$\HH_h$,
in~the sense that
$$ \into K_h(X,Y) f(Y) \,d\mu_h(Y) = f(X), \qquad\forall f\in\HH_h,
\ \forall X\in\Omega.  $$

Our second domain will be the subset $\OmegaN=\{Z\in\CCNN:\;Z^*Z=ZZ^*\}$ of all
normal matrices in~$\CCNN$. By~the Spectral Theorem, any $Z\in\OmegaN$ can be
written in the form
\begin{equation}  Z= U^* D U,  \label{tag:UDU}  \end{equation}
with $U\in\UN$ unitary and $D$ diagonal; $D$~is determined by $Z$ uniquely up
to permutation of the diagonal elements, and if the latter are all distinct and
their order has been fixed in some way, then $U$ is unique up to left
multiplication by a diagonal matrix with unimodular elements. Consequently,
there exists a unique measure $d\mu_h(Z)$ on $\OmegaN$ such that
$$ \intoN f(Z) \,d\mu_h(Z) = (\pi h)^{-N} \int_\UN \int_\CCN
f(U^* DU) \, e^{-\|D\|^2/h} \,dU \, dD  \qquad \forall f, $$
where $dU$ is the normalized Haar measure on~$\UN$, and $dD$ is the Lebesgue
measure on~$\CCN$, where we are identifying the diagonal matrix $D=\diag(d_1,
\dots,d_N)$ with the vector $d=(d_1,\dots,d_N)\in\CCN$,
and $\|D\|^2=\|d\|^2:=|d_1|^2+\dots+|d_N|^2$.
Again, one easily checks \cite{AEG} that
\begin{equation}  \intoN Z^{*j} Z^k \, d\mu_h(Z) = \delta_{jk} k! h^k I,
\label{tag:UOG}  \end{equation}
so~that the elements (\ref{tag:ZK}) are orthogonal also in
$L^2_\CCN(\OmegaN,d\mu_h)$, and we let $\HH_h$ be the subspace spanned by them.
The reproducing kernel is then given~by
\begin{equation}  K_h(X,Y)= \sum_{k=0}^\infty \frac{X^k Y^{*k}}{k! h^k}
\label{tag:KN}  \end{equation}
(with the series converging for all $X,Y\in\CCNN$), in~the sense that
$$ \intoN K_h(X,Y) f(Y) \,d\mu_h(Y) = f(X), \qquad \forall f\in\HH_h,
 \ \forall X\in\OmegaN.  $$

\begin{remark*} At~first sight, the most natural candidate for the
vector-valued space $\HH$ would seem to be the subspace $L^2_{\hol,\CCN}
(\Omega,d\mu)$ of \emph{all} holomorphic functions in $L^2_\CCN(\Omega,d\mu)$
(i.e.~of all square-integrable $\CCN$-valued functions which depend
holomorphically on the coordinates $z_{11},\dots,z_{NN}$ of the point
$z\in\Omega$). However, this choice turns out to be too simple-minded:
the~reproducing kernel is then just $k(x,y)I$, where $k(x,y)$ is the
reproducing kernel of the ordinary (scalar-valued) space $L^2_\hol
(\Omega,d\mu)$; the Toeplitz operator $T_\phi$ (to~be introduced in the next
section) is~just the $N\times N$ matrix $[T_{\phi_{jk}}]_{j,k=1}^N$ of Toeplitz
operators on~$L^2_\hol(\Omega,d\mu)$; and the Berezin transforms (also to be
introduced in the next section) are just $\wt{T_\phi}=[\wt{T_{\phi_{jk}}}]_
{j,k=1}^N$ and $\wt{T_\phi T_\psi}=[\sum_{l=1}^N \wt{T_{\phi_{jl}}
T_{\psi_{lk}}}]_{j,k=1}^N$. Thus, for instance, for the spaces (\ref{tag:SB})
with $n=1$ (i.e.~on~$\Omega=\CC$), $\|\Th_\phi\Th_\psi- \Th_{\phi\psi}\|\to0$
as $h\to0$, while
$$ \|\tfrac{2\pi}{ih}[\Th_\phi,\Th_\psi] - \Th_{\bbk{\phi,\psi}} \| \to 0,  $$
where
$$ \bbk{\phi,\psi} := \partial\phi \,\cdot\, \dbar\psi
- \partial\psi \,\cdot\,\dbar\phi  $$
(here $\partial,\dbar$ are applied individually to each element of a matrix,
and the dot stands for matrix multiplication). From the physical point of view,
this ``matrix-valued Poisson bracket'' seems to be a rather doubtful object,
indicating that the spaces $L^2_{\hol,\CCN}$ are probably not the right route
to~take.  \end{remark*}

\section{The Berezin transform}\label{BTFM}
Let, quite generally, $\HH$~be a reproducing kernel subspace of $\CCN$-valued
functions in $L^2_\CCN(\Omega,d\mu)$, for some domain $\Omega$ and
measure~$\mu$ on~it, with reproducing kernel~$K$; thus $K$ is a $\CCNN$-valued
function on~$\Omega$ and
\begin{equation}  f(X) = \into K(X,Y) f(Y) \,d\mu(Y) \qquad \forall
X\in\Omega,\ f\in\HH.   \label{tag:RP}  \end{equation}
In~particular, for any $\chi\in\CCN$, the functions
$$ K_{Y,\chi}(X):=K(X,Y)\chi  $$
belong to~$\HH$, and
\begin{equation}  \spr{f,K_{Y,\chi}}_\HH \equiv \into K_{Y,\chi}^* f\,d\mu =
\chi^*f(Y). \label{tag:RQ}  \end{equation}
See~\cite{AAG} for more information on such spaces and their reproducing
kernels.

Let $T$ be an arbitrary bounded linear operator on~$\HH$. For any fixed
$X\in\Omega$, the expression $\spr{TK_{X,\chi},K_{X,\chi'}}$ is evidently
linear in $\chi$ and $\chi^{\prime*}$; thus there exists a unique $N\times N$
matrix $\wtt T(X)$ such that
$$ \chi^{\prime*}\wtt T (X) \chi = \spr{TK_{X,\chi},K_{X,\chi'}}.  $$
We~define the \emph{Berezin transform} $\wt T$ of $T$~by
$$ \wt T(X) := K(X,X)^{-1/2} \wtt T(X) K(X,X)^{-1/2}.  $$
That is,
$$ \chi^{\prime*} \wt T(X) \chi = \spr{T K_{X,K(X,X)^{-1/2}\chi},
K_{X,K(X,X)^{-1/2}\chi'}}_\HH.  $$
At~first sight, this definition may seem a little \emph{ad~hoc;} the reason
behind it is that this seems to be the only way to make the following statements
true.

\begin{proposition}\label{thm:PRA} $\wt T$ is a $\CCNN$-valued function on
$\Omega$ satisfying
\begin{enumerate}
\item[\rm(i)] $\wt{T^*}=(\wt T)^*$;
\item[\rm(ii)] if $\phi$ is a matrix-valued function on $\Omega$ such that
$\phi K_{X,\chi}\in\HH$ for all $X\in\Omega$ and $\chi\in\CCN$, then
$$ \wt M_\phi (X) = K(X,X)^{-1/2} \phi(X) K(X,X)^{1/2}  $$
where $M_\phi f:=\phi f$;
\item[\rm(iii)] in~particular, $\wt I(X) = I$ $\forall X\in\Omega$;
\item[\rm(iv)] $\|\wt T(X)\|_{\CCN\to\CCN} \le \|T\|_{\HH\to\HH}$, $\forall
X\in\Omega$. \end{enumerate}
If~the elements of $\HH$ are holomorphic functions, then also
\begin{enumerate}
\item[\rm(v)] $\wt T(X)=0$ $\forall X$ only if $T=0$.
\end{enumerate}   \end{proposition}

\begin{proof} (i)~is immediate from the definition, while (ii) follows from
the reproducing property~(\ref{tag:RP}), and (iii) is a trivial special case
of~(ii). To~prove~(iv), observe that, by~(\ref{tag:RQ}),
$$ \|K_{X,\chi}\|_\HH^2 = \spr{K_{X,\chi},K_{X,\chi}}_\HH = \chi^*
K_{X,\chi}(X) = \chi^*K(X,X)\chi = \|K(X,X)^{1/2}\chi\|^2_\CCN.  $$
Thus, for any $\chi,\chi'\in\CCN$,
\begin{align*}
|\chi^{\prime*}\wt T(X)\chi| &= |\spr{TK_{K(X,X)^{-1/2}\chi},
K_{K(X,X)^{-1/2}\chi'}}| \\
&\le \|T\| \; \|K_{X,K(X,X)^{-1/2}\chi}\| \; \|K_{K(X,X)^{-1/2}\chi'}\|  \\
&= \|T\| \; \|\chi\|_\CCN \; \|\chi'\|_\CCN,  \end{align*}
and the assertion follows.

Finally, for~(v), note that $\wt T\equiv0$ implies $\wtt T\equiv0$,~i.e.
\begin{equation}  \spr{TK_{Y,\chi'},K_{X,\chi}} =0 \qquad\forall \chi,\chi'
\label{tag:PET}  \end{equation}
whenever $X=Y$. If~$\HH$ consists of holomorphic functions, then $K(X,Y)$ is
holomorphic in $X$ and conjugate-holomorphic in~$Y$; thus the left-hand side of
(\ref{tag:PET}) is holomorphic in $X$ and conjugate-holomorphic in~$Y$. It~is
well known that such functions are uniquely determined by their restriction to
the diagonal $X=Y$; consequently, (\ref{tag:PET})~holds for all $X,Y$,~i.e.
$$ \chi^* (TK_{Y,\chi'})(X) =0 \qquad \forall \chi,\chi' \ \forall X,Y.  $$
Hence $TK_{Y,\chi'}=0$ for all $Y$ and $\chi'$, and thus for any $f\in\HH$
$$ \chi^{\prime*}(T^*f)(Y) = \spr{T^*f,K_{Y,\chi'}}=\spr{f,TK_{Y,\chi'}}=0, $$
i.e.~$T^*f=0$. Thus $T^*=0$ and $T=0$.    \end{proof}

In~analogy with the scalar-valued situation, we~next define for any $\phi\in
C^\infty_\CCNN(\Omega)$ the Toeplitz operator $T_\phi$ on $\HH$ by the recipe
$$ T_\phi f(X) := \into K(X,Y) \phi(Y) f(Y) \,d\mu(Y),  $$
or,~equivalently,
$$ T_\phi f = \bP(\phi f),  $$
where $\bP:L^2_\CCN(\Omega,d\mu)\to\HH$ is the orthogonal projection. Note that
the last formula implies that $\|T_\phi\|_{\HH\to\HH} \le \|\phi\|_\infty:=
\sup_{X\in\Omega}\|\phi(X)\|_{\CCN\to\CCN}$; in~particular,
by~Proposition~\ref{thm:PRA},
also $\|\wt T_\phi(X)\|_{\CCN\to\CCN}\le\|\phi\|_\infty$.

\begin{proposition}\label{thm:PRB} The following formulae hold:
\begin{align*}
\wt T_\phi(X) &= K(X,X)^{-1/2} \cdot \into K(X,Y)\phi(Y)K(Y,X)\,d\mu(Y) \cdot
K(X,X)^{-1/2} ;  \\
\wt{T_\phi T_\psi}(X) &= K(X,X)^{-1/2}  \\
& \qquad\qquad \cdot \into\into K(X,Y)\phi(Y) K(Y,Z)
\psi(Z) K(Z,X) \, d\mu(Y) \,d\mu(Z)  \\
& \qquad\qquad\qquad\qquad  \cdot K(X,X)^{-1/2}.   \end{align*}
In~particular, if $\phi$ is a multiplier of~$\HH$ $($i.e.~$\phi f\in\HH$
whenever $f\in\HH)$, then
$$ \wt T_\phi (X) = K(X,X)^{-1/2} \phi(X) K(X,X)^{1/2};  $$
and, similarly, when $\phi$ and $\psi^*$ are multipliers of~$\HH$, then
$$ \wt{T_\phi T_\psi}(X) = K(X,X)^{-1/2}\phi(X)K(X,X)\psi(X)K(X,X)^{-1/2}.  $$
\end{proposition}

\section{Bad behaviour: all matrices}\label{BBFM}
We~now exhibit an example of some pathological phenomena, showing in particular
that the straightforward generalizations of the expansions (\ref{tag:AL})
and~(\ref{tag:AM}) cannot hold: first, apart from the integer powers of~$h$,
we~will see that also $\sqrt h$ enters the picture; and second, instead of
evaluations at $X$ we get also contributions from other points.
This applies to the case of the full matrix domain $\Omega=\CCNN$; for the
normal matrices~$\OmegaN$, we~will show in the next section that at least the
second of these pathologies still remains.

\begin{theorem}\label{thm:THA} Consider the full matrix domain $\Omega=\CCNN$
with $N=2$. Let $\HH_h$ be the spaces from Section~\ref{SPCS}, with reproducing
kernels $K_h$ given by~$(\ref{tag:KH})$. Let $X$ be the matrix
$$ X=\begin{pmatrix} 0 & 1 \\ 0 & 0 \end{pmatrix} .  $$
Then:
\begin{enumerate}
\item[\rm(i)] $\displaystyle \lim_{h\to0} \wt{\Th_\phi}(X) = \begin{pmatrix}
\frac{\phi_{11}(0)+\phi_{22}(0)}2 & 0 \\ 0 & \phi_{22}(0) \end{pmatrix}$.
\item[] Note that the matrix on the right-hand side does not depend on the
value of $\phi$ at~$X$, but rather on its value at~$0$.
\item[\rm(ii)] If~$\phi(Y)=\sqrt2 y_{22} I$, where $y_{22}$ denotes the
$(2,2)$-entry of~$Y$, then
$$ \wt{\Th_\phi}(X) = X\sqrt h +O(h) \quad\text{as }h\to0.  $$
\end{enumerate}  \end{theorem}

\begin{proof} Since $X^2=0$, the series (\ref{tag:KH}) for $K_h(X,Y)$ becomes
simply
\begin{equation}  K_h(X,Y) = I+ \frac{XY^*}{c_1 h} = I+\frac{XY^*}{2h}.
\label{tag:VC}  \end{equation}
Thus
\begin{align}
& \wtt{\Th_\phi}(X) = \into K_h(X,Y)\phi(Y) K_h(Y,X)\,d\mu_h(Y) \nonumber \\
& \qquad\qquad = \into\bigg[\phi(Y) + \frac{XY^*\phi(Y)}{2h} +
\frac{\phi(Y)YX^*}{2h} + \frac{XY^*\phi(Y)YX^*}{4h^2} \bigg] \,d\mu_h(Y).
\label{tag:VA}   \end{align}
On~the other hand, the change of variable $Z=Y/\sqrt h$ and Taylor's expansion
imply that for any $C^\infty$ function $f$ on~$\Omega$,
\begin{equation}  \begin{aligned}
\into f(Y) \,d\mu_h(Y) &= \into f(\sqrt h Z)\,d\mu_1(Z)  \\
&= \sum_{j=0}^\infty \frac1{j!} \Delta^j f(0) \, h^j,  \end{aligned}
\label{tag:VB}  \end{equation}
where $\Delta=\sum_{j,k=1}^2 \partial^2/\partial z_{jk}\partial\oz_{jk}$ is the
Laplacian on~$\CC^{2\times 2}$. Applying this to~(\ref{tag:VA}), we~therefore
get
\begin{align*}
\wtt{\Th_\phi}(X) &= \phi(0) + O(h) \\
&\quad + \frac{X[\Delta(Y^*\phi(Y))(0)+O(h)]}2 \\
&\quad + \frac{[\Delta(\phi(Y)Y)(0)+O(h)]X^*}2 \\
&\quad + X\Big[\frac{\Delta(Y^*\phi(Y)Y)(0)}{4h}+O(1)\Big]X^*.  \end{align*}
Now $\wt{\Th_\phi}(X)=K(X,X)^{-1/2}\wtt{\Th_\phi}(X)K(X,X)^{-1/2}$; note that
$$ K_h(X,X)^{-1/2} = \begin{pmatrix} \sqrt{\frac{2h}{2h+1}} &0
\\0&1\end{pmatrix},  $$
and thus, for any matrix $A=\begin{pmatrix}
a_{11}&a_{12}\\a_{21}&a_{22}\end{pmatrix}$,
\begin{equation}  \begin{aligned}
K(X,X)^{-1/2}AK(X,X)^{-1/2}
&= \begin{pmatrix} \frac{2h}{2h+1} a_{11} & \sqrt{\frac{2h}{2h+1}} a_{12} \\
   \sqrt{\frac{2h}{2h+1}} a_{21} & a_{22} \end{pmatrix},  \\
K(X,X)^{-1/2}XAK(X,X)^{-1/2}
&= \begin{pmatrix} \frac{2h}{2h+1} a_{21} & \sqrt{\frac{2h}{2h+1}} a_{22} \\
    0 & 0 \end{pmatrix},  \\
K(X,X)^{-1/2}AX^*K(X,X)^{-1/2}
&= \begin{pmatrix} \frac{2h}{2h+1} a_{12} & 0 \\
   \sqrt{\frac{2h}{2h+1}} a_{22} & 0 \end{pmatrix},  \\
K(X,X)^{-1/2}XAX^*K(X,X)^{-1/2}
&= \begin{pmatrix} \frac{2h}{2h+1} a_{22} & 0 \\ 0 & 0 \end{pmatrix} .
\end{aligned}  \label{tag:KKKK}   \end{equation}
Consequently,
\begin{align*}
\wt{\Th_\phi}(X)
&= \begin{pmatrix} 0 & \phi_{12}(0)\sqrt{2h} \\ \phi_{21}(0) \sqrt{2h} &
\phi_{22}(0) \end{pmatrix} + O(h)  \\
&\quad + \begin{pmatrix} 0 & \frac{\sqrt{2h}}2 \Delta[Y^*\phi(Y)]_{22}(0) \\
0 & 0 \end{pmatrix} + O(h) \\
&\quad + \begin{pmatrix} 0 & 0 \\ \frac{\sqrt{2h}}2 \Delta[\phi(Y)Y]_{22}(0) &
0 \end{pmatrix} + O(h) \\
&\quad + \begin{pmatrix} \frac12 \Delta[Y^*\phi(Y)Y]_{22}(0) & 0 \\ 0 & 0
\end{pmatrix} + O(h) .   \end{align*}
Note that, by~a simple calculation,
\begin{align}
\Delta[Y^*\phi(Y)]_{ab}(0) &= \sum_{k=1}^2 \frac{\partial\phi_{kb}}
{\partial y_{ka}}(0), \label{tag:SICA}  \\
\Delta[\phi(Y)Y]_{ab}(0) &= \sum_{k=1}^2 \frac{\partial\phi_{ak}}
{\partial\oy_{kb}}(0), \nonumber \\
\Delta[Y^*\phi(Y)Y]_{ab}(0) &= \delta_{ab} \cdot \Tr\phi(0).  \nonumber
\end{align}
Letting $h\searrow0$, (i)~therefore follows immediately. For
$\phi(Y)=y_{22}\sqrt2 I$, only the second term in the last formula for
$\wt{\Th_\phi}(X)$ gives a nonzero contribution, equal~to
$$ \wt{\Th_\phi}(X) = \begin{pmatrix} 0 & \sqrt h \,\Delta[y_{22}Y^*]_{22}(0)
\\ 0 & 0 \end{pmatrix} + O(h) = \sqrt h \, X + O(h),  $$
as $\Delta[y_{22}Y^*]_{22}(0)=1$ by~(\ref{tag:SICA}). This settles~(ii).
 \end{proof}

\begin{remark*} We~pause to observe that applying (\ref{tag:VB}) to
$f(Y)=Y^{*k}Y^k$ and comparing with~(\ref{tag:ZZ}) shows that
\begin{equation} \Delta^k(Y^{*k}Y^k)(0) = k! c_k I.  \label{tag:NICE}
\end{equation}
This gives, conceivably, a~way of evaluating the numbers $c_k$ without
recourse to random matrix theory, and can also  be used to show that the $c_k$
have an interesting combinatorial meaning. Namely, expanding $\Delta^k$ and
$Y^{*k}Y^k$ yields
\begin{multline*} \Delta^k(Y^{*k}Y^k) = \sum_{i_1,j_1,\dots,i_k,j_k=1}^N
\partial_{i_1 j_1}\dbar_{i_1 j_1} \dots \partial_{i_k j_k}\dbar_{i_k j_k} \\
\sum_{a_1,\dots,a_{2k-1}=1}^N \oy_{a_1 a}\oy_{a_2 a_1}\dots\oy_{a_k a_{k-1}}
y_{a_k a_{k+1}} y_{a_{k+1}a_{k+2}} \dots y_{a_{2k-1}b},  \end{multline*}
where, for the sake of brevity, we~temporarily write $\partial_{ij}$ for
$\partial^2/\partial y_{ij}$, and similarly for~$\dbar_{ij}$.
Clearly a nonzero contribution only occurs if to each $y$ there is applied
precisely one $\partial$, and to each $\oy$ precisely one~$\dbar$. (We~also
see that the result will be independent of~$Y$, i.e.~a~constant.) Thus
\begin{multline*} \Delta^k(Y^{*k}Y^k) = \sum_{i_1,j_1,\dots,i_k,j_k=1}^N
\sum_{a_1,\dots,a_{2k-1}=1}^N \sum_{\sigma,\tau\in\mathfrak S_k}
\dbar_{i_{\tau(1)}j_{\tau(1)}} \oy_{a_1 a} \dots
\dbar_{i_{\tau(k)}j_{\tau(k)}} \oy_{a_k a_{k-1}} \\
\vphantom{\Big\|} \cdot \partial_{i_{\sigma(1)}j_{\sigma(1)}} y_{a_k a_{k+1}}
\dots \partial_{i_{\sigma(k)}j_{\sigma(k)}} y_{a_{2k-1}b}.  \end{multline*}
Changing the order of summations and using the fact that $\partial_{ij}y_{kl}=
\delta_{ik}\delta_{jl}$, this becomes
\begin{multline*} \Delta^k(Y^{*k}Y^k) = \sum_{\sigma,\tau\in\mathfrak S_k}
\sum_{i_1,j_1,\dots,i_k,j_k=1}^N \sum_{a_1,\dots,a_{2k-1}=1}^N
\delta_{i_{\tau(1)} a_1} \delta_{j_{\tau(1)} a} \cdot \dots
\delta_{i_{\tau(k)} a_k} \delta_{j_{\tau(k)} a_{k-1}} \\ \vphantom{\Big\|}
\cdot \delta_{i_{\sigma(1)} a_k} \delta_{j_{\sigma(1)} a_{k+1}} \cdot \dots
\delta_{i_{\sigma(k)} a_{2k-1}} \delta_{j_{\sigma(k)} b}.   \end{multline*}
Replacing $\mathbf i=(i_1,\dots,i_k),\mathbf j=(j_1,\dots,j_k)$ by $\mathbf i
\circ\tau^{-1},\mathbf j\circ\tau^{-1}$ and setting $\mu=\tau^{-1}\sigma$,
we~thus get
\begin{multline*} \Delta^k(Y^{*k}Y^k) = k! \sum_{\mu\in\mathfrak S_k}
\sum_{\mathbf i,\mathbf j}
\delta_{j_1 a} \delta_{j_2 i_1} \dots \delta_{j_k i_{k-1}} \\
\vphantom{\Big\|} \cdot \delta_{i_{\mu(1)} i_k} \delta_{i_{\mu(2)} j_{\mu(1)}}
\dots \delta_{i_{\mu(k)} j_{\mu(k-1)}} \delta_{j_{\mu(k)} b}.  \end{multline*}
In~other words, $[\Delta^k(Y^{*k}Y^k)]_{ab}/k!=c_k\delta_{ab}$ is the constant
equal to the number of triples $(\mu,\mathbf i,\mathbf j)$, where
$\mu\in\mathfrak S_k$ and $\mathbf i,\mathbf j\in\{1,\dots,N\}^k$, such that
$$ \mathbf j=(a,i_1,\dots,i_{k-1}), \quad \mathbf j\circ\mu =
(i_{\mu(2)},\dots,i_{\mu(k)},b), \quad  \text{and} \quad i_k=i_{\mu(1)}.  $$
It~is evident that, indeed, this number is zero for $a\neq b$ (since the
sequences $(a,i_1,\dots,i_{k-1},i_k)$ and $(i_{\mu(1)},i_{\mu(2)},\dots,
i_{\mu(k)},b)$ must be permutations of each other), while for $a=b$ it is
independent of~$a$. It~is also clear that $c_k$ is always an integer, a~fact
definitely not apparent from~(\ref{tag:CK}).   \qed   \end{remark*}

We~conclude by giving a formula analogous to part~(i) of the last theorem also
for $\wt{\Th_\phi\Th_\psi}(X)$. It~shows, in~particular, that $\wt{\Th_\phi
\Th_\psi}(X)-\wt{\Th_{\phi\psi}}(X)$ need not tend to~0 in general as $h\to0$,
but does so for scalar-valued $\phi$ and~$\psi$.

\begin{theorem}\label{thm:THB} Under the hypotheses of Theorem~\ref{thm:THA},
$$ \lim_{h\to0} \wt{\Th_\phi\Th_\psi}(X) = \begin{pmatrix} \frac{\Tr\phi(0)}2
\cdot \frac{\Tr\psi(0)}2 & 0 \\ 0 & (\phi\psi)_{22}(0) \end{pmatrix}.   $$
\end{theorem}

\begin{proof} Using the formula for $\wt{\Th_\phi\Th_\psi}$ from
Proposition~{\ref{thm:PRB}} and~(\ref{tag:VC}), we~have
$$ \wtt{\Th_\phi\Th_\psi}(X) = \into\into \Big(I+\frac{XY^*}{2h}\Big) \phi(Y)
K_h(Y,Z) \psi(Z) \Big(I+\frac{ZX^*}{2h}\Big) \,d\mu_h(Y) \,d\mu_h(Z).   $$
Making again the change of variable $Y\mapsto Y\sqrt h$, $Z\mapsto Z\sqrt h$,
the double integral becomes
$$ \into\into \Big(I+\frac{XY^*\sqrt h}{2h}\Big) \phi(\sqrt h Y) K_1(Y,Z)
\psi(\sqrt h Z) \Big(I+\frac{Z\sqrt h X^*}{2h}\Big) \,d\mu_1(Y) \,d\mu_1(Z). $$
The rest of the proof proceeds in the same way as in Theorem~\ref{thm:THA},
using instead of~(\ref{tag:VB}) the expansion from the following lemma (applied
also to $Y^*\phi(Y)$ and $\psi(Z)Z$ in  place of $\phi(Y)$ and $\psi(Z)$,
respectively), the~fact that
$$ \bbr{Y^*\phi(Y),\psi(Z)Z}_{22} = \frac14\Tr\phi(0)\Tr\psi(0),   $$
and the relations~(\ref{tag:KKKK}). We~leave the details to the reader.
\end{proof}

\begin{lemma}\label{thm:LEM} For $\phi,\psi\in C^\infty_\CCNN(\Omega)$,
\begin{align*}
& \into\into \phi(\sqrt h Y)K_1(Y,Z)\psi(\sqrt h Z) \,d\mu_1(Y) \,d\mu_1(Z) \\
&\qquad = \phi(0)\psi(0) +h \big[\Delta\phi(0)\cdot\psi(0) +
\phi(0)\cdot\Delta\psi(0) + \bbr{\phi,\psi}(0) \big] + O(h^2),  \end{align*}
as~$h\to0$, where
$$ \bbr{\phi,\psi} := \frac12 \sum_{i,j,k}
\frac{\partial\phi}{\partial\oy_{ij}} \, E_{ik} \,
\frac{\partial\psi}{\partial z_{kj}},   $$
where $E_{ik}$ is the matrix $[E_{ik}]_{ab}=\delta_{ia}\delta_{kb}$.
\end{lemma}

\begin{proof} Using the Taylor expansions for $\phi$ and~$\psi$, we~see that
the integral asymptotically equals
$$ \sum_{\alpha,\beta,\gamma,\delta}
h^{\frac{|\alpha|+|\beta|+|\gamma|+|\delta|}2}
\frac{\partial^\alpha\dbar^\beta\phi(0)}{\alpha!\beta!}
\into\into y^\alpha\oy^\beta K_1(Y,Z) z^\gamma\oz^\delta \,d\mu_1(Y)\,d\mu_1(Z)
\;\frac{\partial^\gamma\dbar^\delta\psi(0)}{\gamma!\delta!} .  $$
(The summation extends over all multiindices $\alpha,\beta,\gamma,\delta$.)
Note that the kernel satisfies $K_1(Y,Z)=K_1(\epsilon Y,\epsilon Z)$ for any
$\epsilon\in\CC$ of modulus~one; hence the last integral vanishes unless
$|\alpha|+|\gamma|=|\beta|+|\delta|$. Thus the coefficients at half-integer
powers of~$h$ in fact vanish. The coefficient at~$h^0$ is clearly
$\phi(0)\psi(0)$, since
$$ \into \into K_1(Y,Z)\,d\mu_1(Y) \,d\mu_1(Z) = \into I \,d\mu_1(Y) = I  $$
by the reproducing property of $K_1$ and~(\ref{tag:ZZ}). For the coefficient
at~$h^1$, the only nonzero contributions, by~virtue of the last observation,
come from $|\alpha|=|\beta|=1$, or $|\gamma|=|\delta|=1$, or
$|\alpha|=|\delta|=1$, or $|\beta|=|\gamma|=1$ (i.e.~from $y\oy$, $z\oz$,
$y\oz$, or $\oy z$). Since
$$ \into\into y_{ij} \oy_{kl} K_1(Y,Z) \,d\mu_1(Y)\,d\mu_1(Z) =
\into y_{ij}\oy_{kl} I \,d\mu_1(Y) = \delta_{ik} \delta_{jl}I  $$
(and similarly for $z_{ij}\oz_{kl}$), the first two possibilities contribute
$$ \sum_{i,j,k,l} \frac{\partial^2\phi(0)}{\partial y_{ij}\partial\oy_{kl}}
\;\delta_{ik}\delta_{jl}I\; \psi(0) = \Delta\phi(0) \cdot \psi(0)  $$
and $\phi(0)\cdot\Delta\psi(0)$, respectively. For the $y\oz$ possibility, the
corresponding integral vanishes, since the integrand is a function holomorphic
in the entries of $Y$ and $Z^*$ and vanishing at the origin. Finally, for the
last possibility $\oy z$ we use the series (\ref{tag:KH}) to split the
integral~as
\begin{align}
& \into\into \oy_{ij} z_{kl} K_1(Y,Z) \,d\mu_1(Y) \,d\mu_1(Z)   \label{tag:YZ}
\\ &\qquad\qquad = \sum_{m=0}^\infty \frac1{c_m}
\Big(\into \oy_{ij} Y^m \,d\mu_1(Y)\Big)
\Big(\into z_{kl} Z^{*m} \,d\mu_1(Z) \Big).   \nonumber  \end{align}
Using again the invariance of $d\mu_1$ under the change of variable
$Y\mapsto\epsilon Y$, we~see that we only get nonzero contribution for $m=1$.
In~that case,
$$ \Big[\into \oy_{ij} Y\,d\mu_1(Y)\Big]_{ab} = \into \oy_{ij} y_{ab} \,
d\mu_1(Y) = \delta_{ia}\delta_{jb} = [E_{ij}]_{ab},  $$
and similarly for the $Z$ integral. Thus the integral (\ref{tag:YZ}) equals
$$ \frac1{c_1} E_{ij} E_{lk} = \frac12 \delta_{jl} E_{ik} ,  $$
and the total contribution from the $\oy z$ possibility~is
$$ \sum_{i,j,k,l} \frac{\partial\phi}{\partial\oy_{ij}}(0) \;
\frac{\delta_{jl}}2 \, E_{ik} \; \frac{\partial\psi}{\partial z_{kl}}(0)
= \bbr{\phi,\psi}(0),  $$
which concludes the proof of the lemma.     \end{proof}

\section{Bad behaviour: normal matrices}\label{BBNM}
The matrix $X=\begin{pmatrix} 0&1\\0&0\end{pmatrix}$ featuring in the last
section was not normal; this tempts one to hope that things might perhaps still
work out fine for the domain $\OmegaN$ of all normal matrices. We~show that
even in this case, unfortunately, the non-local behaviour described above still
persists.

\begin{theorem}\label{thm:THC} Consider the domain $\OmegaN$ of all normal
$N\times N$ matrices, with $N=2$. Let $\HH_h$ be the spaces from
Section~\ref{SPCS}, with reproducing kernels $K_h$ given by~$(\ref{tag:KN})$.
Let $X$ be the matrix
$$ X=\begin{pmatrix} 1&0 \\ 0&0 \end{pmatrix}  $$
of~the projection onto the first coordinate. Then
\begin{align*}
[\wt{\Th_\phi}(X)]_{11} &= [\wt{\Th_\phi}(I)]_{11} ,  \\
[\wt{\Th_\phi}(X)]_{22} &= [\wt{\Th_\phi}(0)]_{22} .  \end{align*}
Consequently, the asymptotic expansion $(\ref{tag:AL})$ cannot hold.
\end{theorem}

\begin{proof} As~$X$ is a projection, we~have $X^j=X$ $\forall j\ge1$; thus
$$ K_h(X,Y) = \sum_{j=0}^\infty \frac{X^jY^{*j}}{j! h^j} = I+X\sum_{j=1}^\infty
\frac{Y^{*j}}{j!h^j} = (I-X)+X K_h(I,Y).  $$
Thus
\begin{align*}
\wtt{\Th_\phi}(X) &= \intoN K_h(X,Y) \phi(Y) K_h(Y,X) \,d\mu_h(Y) \\
&= X\cdot \intoN K_h(I,Y)\phi(Y) K_h(Y,I)\,d\mu_h(Y) \cdot X  \\
&\qquad + X\cdot \intoN K_h(I,Y)\phi(Y) \,d\mu_h(Y) \cdot (I-X)  \\
&\qquad + (I-X)\cdot \intoN \phi(Y) K_h(Y,I)\,d\mu_h(Y) \cdot X  \\
&\qquad + (I-X)\cdot \intoN \phi(Y) \,d\mu_h(Y) \cdot (I-X).  \end{align*}
But for any matrix $A=\begin{pmatrix} a_{11}&a_{12}\\a_{21}&a_{22}
\end{pmatrix}$, we~have $XAX=\begin{pmatrix} a_{11}&0\\0&0\end{pmatrix}$,
$XA(I-X)=\begin{pmatrix} 0&a_{12} \\0&0\end{pmatrix}$, etc.; hence
$$ [\wtt{\Th_\phi}(X)]_{11}
= \Big[\intoN K_h(I,Y)\phi(Y) K_h(Y,I)\,d\mu_h(Y)\Big]_{11}
= [\wtt{\Th_\phi}(I)]_{11},    $$
and similarly, since $K_h(0,Y)=I$,
$$ [\wtt{\Th_\phi}(X)]_{22} = \Big[\intoN \phi(Y) \,d\mu_h(Y)\Big]_{22}
= [\wtt{\Th_\phi}(0)]_{22} .   $$
Finally, since $K_h(X,X)=\begin{pmatrix} e^{1/h} &0\\0&1\end{pmatrix}$,
$K_h(I,I)= e^{1/h} I$, and $K_h(0,0)=I$, the assertion about $\wt{\Th_\phi}(X)
=K(X,X)^{-1/2}\wtt{\Th_\phi}(X)K(X,X)^{-1/2}$ follows.      \end{proof}

For completeness, we~also state the analog of Theorem~\ref{thm:THB}, which
shows, among others, that the expansion~(\ref{tag:AM}) cannot hold. Its~proof
is the same as  for Theorem~\ref{thm:THC}.

\begin{theorem}\label{thm:THD} In~the situation of the preceding theorem,
\begin{align*}
[\wt{\Th_\phi\Th_\psi}(X)]_{11} &= [\wt{\Th_\phi\Th_\psi}(I)]_{11},  \\
[\wt{\Th_\phi\Th_\psi}(X)]_{22} &= [\wt{\Th_\phi\Th_\psi}(0)]_{22}.
\end{align*}    \end{theorem}

\section{An application of stationary phase}\label{STPH}
In~this section we finally start exhibiting also results in the positive
direction, namely, by~using the stationary phase method we establish the
existence of~an (albeit non-local) semiclassical asymptotic expansion for
$\wt{\Th_\phi}$ for the case of the normal matrices.

Recall that the stationary phase (WJKB) method tells us that if $S$, $\phi$ are
smooth complex-valued functions on some domain in~$\CC^n$, such that $S$ has a
unique critical point~$x_0$ (i.e.~$S'(x_0)=0$), which is nondegenerate
(i.e.~$\det S''(x_0))\neq0$) and is a global maximum for $\RE S$, and $\phi$ is
compactly supported, then the integral
\begin{equation}  h^{-n} \int \phi(x) \; e^{S(x)/h} \, dx  \label{tag:HS}
\end{equation}
has an asymptotic expansion
\begin{equation}  e^{S(x_0)/h} \sum_{j=0}^\infty h^j \; \LL_j\phi(x_0)
 \qquad\text{as } h\to 0,  \label{tag:SPH}   \end{equation}
with some differential operators $\LL_j$ whose coefficients are given by
universal expressions in $S$ and its partial derivatives. See e.g.~\cite{Hrm},
Section~7.7. The~hypothesis of the compact support of $\phi$ can be replaced by
the requirement that the integral (\ref{tag:HS}) exist for some $h=h_0>0$, and
that the maximum of $\RE S$ at $x_0$ strictly dominate also the values of
$\RE S$ at the boundary or at  infinity, in~the sense that $\RE S(x_n)\to
\RE S(x_0)\implies x_n\to x_0$.

On~the other hand, if the global maximum of $\RE S$ is not a critical point,
then (\ref{tag:HS}) decays faster than any power of $h$ as~$h\searrow0$.

The~formulas for the operators $\LL_j$ are fairly complicated in general, but
fortunately become quite explicit if the phase function $S$ is quadratic (which
will be the only case we will need). Namely, assume that
$$ S(x) = -\spr{Q(x-x_0),x-x_0}_{\CC^n}  $$
for some matrix $Q$ with positive real part. Then $x_0$ is a unique critical
point of~$S$, is~nondegenerate,~and
\begin{equation}  \LL_j = \frac1{j!} \mathcal Q^j, \qquad\text{where}\quad
 \mathcal Q=-\spr{Q^{-1}\partial,\partial}.  \label{tag:QL}   \end{equation}

Let~us now apply this to the integral defining $\wt{\Th_\phi}(X)$ in the case
of the domain of normal matrices,~viz.
$$ \wt{\Th_\phi}(X) = \intoN K_h(X,X)^{-1/2} K_h(X,Y) \phi(Y) K_h(Y,X)
K_h(X,X)^{-1/2} \, d\mu_h(Y).   $$
Let
\begin{equation}  Y=UDU^*, \qquad X=VCV^*  \label{tag:YX}  \end{equation}
be the spectral decompositions of $Y$ and~$X$, respectively. Observe that owing
to the invariance of the kernels $K_h$ and the measures $\mu_h$ under unitary
transformations, we~have $\wt{\Th_\phi}(X)=V\wt{\Th_{\phi^V}}(C)V^*$, where
$\phi^V(Y):=V^*\phi(VYV^*)V$; thus it suffices to deal with the case of $V=I$,
i.e.~when $X=C=\diag(c_1,\dots,c_N)$ is a diagonal matrix.
From~(\ref{tag:YX}), we~then have
\begin{align*}
[K_h(X,Y)]_{ij} &= \sum_{k=0}^\infty \frac{[C^k U D^{*k} U^*]_{ij}}{k! h^k} \\
&= \sum_{k=0}^\infty \sum_{l=1}^N \frac{c_i^k u_{il}\od_l^k u^*_{lj}}{k!h^k} \\
&= \sum_{l=1}^N e^{c_i\od_l/h} u_{il} \ou_{jl},  \end{align*}
and similarly for $K_h(X,X)$. Thus the matrix entries of $\wt{\Th_\phi}(X)$ are
given~by
$$ \begin{aligned}
& [\wt{\Th_\phi}(X)]_{ab} = (\pi h)^{-N} \int_\UN \int_\CCN
\sum_{j,k,l,m=1}^N e^{-|c_a|^2/2h} e^{c_a\od_l/h} u_{al}\ou_{jl}
\phi_{jk}(UDU^*) \\
& \hskip16em \cdot\; e^{\oc_b d_m/h} \ou_{bm} u_{km} e^{-|c_b|^2/2h}
e^{-\|d\|^2/h} \, dU \, dD.   \end{aligned}  $$
For~simplicity, we~will write $\phi(U;d_1,\dots,d_N)$ instead of $\phi(UDU^*)$.
The~last integral over $D$ is precisely of the form~(\ref{tag:HS}), with phase
function given~by
$$ S(d_1,\dots,d_N) = c_a\od_l+\oc_b d_m-\|d\|^2-\frac{|c_a|^2+|c_b|^2}2.  $$
The critical point condition $S'=0$ amounts~to
$$ c_a\delta_{li}=d_i, \quad\oc_b\delta_{mi}=\od_i,\qquad\forall i=1,\dots,N.$$
It~follows that there is no critical point if $c_a\neq c_b$, or~if $c_a=c_b
\neq0$ and $l\neq m$; while for $c_a=c_b\neq0$ and $l=m$, or~$c_a=c_b=0$ and
$l,m$ arbitrary, there is a unique critical point
$$ d=(0,\dots,0,\overbrace{c_a}^{\text{$l$-th slot}},0,\dots,0) \equiv
c_a\chi_l,  $$
which satisfies the assumptions for the application of the stationary phase
method. The critical value is
$$ S(c_a\chi_l) = |c_a|^2 + |c_a|^2 -|c_a|^2 - \frac{|c_a|^2+|c_a|^2}2 = 0, $$
and the operators $\LL_j$ are equal to $\frac1{j!}\Delta^j$, by~(\ref{tag:QL}).
By~(\ref{tag:SPH}), it~therefore follows that
$[\wt{\Th_\phi}(X)]_{ab}=O(h^\infty)$ for $c_a\neq c_b$, while
\begin{equation}  [\wt{\Th_\phi}(X)]_{ab} \approx \sum_{j,k,l=1}^N \sumr
\;\frac{h^r}{r!} \;
\int_\UN u_{al} \ou_{jl} u_{kl} \ou_{bl} \; (\Delta^r_{(d)}\phi_{jk})
(U;c_a\chi_l) \, dU   \label{tag:ASY}  \end{equation}
as $h\to0$ if $c_a=c_b\neq0$, and
\begin{align}
[\wt{\Th_\phi}(X)]_{ab} &\approx
\sum_{j,k,l,m,=1}^N \sumr \; \frac{h^r}{r!} \; \int_\UN u_{al} \ou_{jl} u_{km}
\ou_{bm} \; (\Delta^r_{(d)} \phi_{jk})(U;0) \, dU  \nonumber  \\
&= \sumr \;\frac{h^r}{r!} \; \int_\UN (\Delta^r_{(d)} \phi_{ab})(U;0) \,dU
\nonumber \\
&= \sumr \;\frac{h^r}{r!} \; (\Delta^r_{(d)} \phi_{ab})(0)
\qquad\text{(as $\phi(U;0)=\phi(0)$ is independent of $U$)}
\label{tag:TSS} \end{align}
as $h\to0$ if $c_a=c_b=0$. Here the subscript at $\Delta$ indicates that it
applies only to the $d$-variables in $\phi(U;d_1,\dots,d_N)$.

Thus the coefficients at each $h^r$ in the asymptotic expansion do not depend
on the jet of $\phi$ at~$X$, but rather on the behaviour of $\phi$ near the
whole orbit $\{UP_a U^*:\;U\in\UN\}$ of the spectral components $P_a:=\diag
(0,\dots,0,c_a,0,\dots,0)$ of~$X$. Also, the off-diagonal entries
asymptotically vanish (i.e.~are $O(h^\infty)$) if $c_a\neq c_b$,
which is quite unexpected.

Observe that setting $c_a=0$ in~(\ref{tag:ASY}) gives
$$ [\wt{\Th_\phi}]_{ab} \approx \sum_{j,k=1}^N \sumr\;\frac{h^r}{r!} \;
\kappa_{jk}  \; \Delta^r_{(d)} \phi_{jk}(0)  $$
where
$$ \kappa_{jk} := \sum_{l=1}^N \; \int_\UN u_{al} \ou_{jl}
u_{kl} \ou_{bl} \,dU.  $$
It~can be shown that
\begin{equation} \kappa_{jk} = \frac{\delta_{aj}\delta_{kb}
+ \delta_{ab}\delta_{kj}}{N+1},   \label{KAPP}  \end{equation}
and in fact,
\begin{equation} \int_\UN u_{al} \ou_{jl} u_{kl} \ou_{bl} \,dU =
\frac{\delta_{aj}\delta_{kb}+\delta_{ab}\delta_{kj}}{N(N+1)}.\label{orthogreln}
\end{equation}
(The above relation can be obtained by an application of standard orthogonality
relations for the matrix elements of irreducible representations of compact
groups --- in this case applied to the irreducible subrepresentation of $U(N)$,
carried by second order symmetric tensors, in~the decomposition of the natural
representation of $U(N)\otimes U(N)$.) Thus we have
$$[\wt{\Th_\phi}]_{ab} \approx \sumr \; \frac{h^r}{r!} \;
\frac{\Delta^r_{(d)}(\phi_{ab}+\delta_{ab}\Tr\phi)(0)}{N+1},$$
which is different from~(\ref{tag:TSS}). Thus we see that, in~general, it~is
not possible to use the~formula~(\ref{tag:ASY}) in both cases $c_a=c_b\neq0$
and $c_a=c_b=0$.

For scalar-valued~$\phi$, (\ref{tag:ASY})~simplifies~to
$$ [\wt{\Th_\phi}(X)]_{ab} \approx \sumr \sum_{l=1}^N \;\frac{h^r}{r!} \;
\int_\UN u_{al} \ou_{bl} \; (\Delta^r_{(d)} \phi) (U;c_a\chi_l) \,dU ;   $$
and if in addition $\phi$ is independent of~$U$, i.e.~$\phi(UYU^*)=\phi(Y)$
$\forall U\in\UN$, then the last integral can be evaluated by Schur's
orthogonality relations, yielding
\begin{equation}  [\wt{\Th_\phi}(X)]_{ab} \approx \delta_{ab} \,\cdot \sumr
\,\frac{h^r}{r!} \; (\Delta^r_{(d)} \phi) (c_a\chi_a).  \label{tag:TDE}
\end{equation}
In~the general case, however, it~does not seem that (\ref{tag:ASY}) can be
simplified in any~way.

In~the same manner, one~can also prove the following formula for the
asymptotics of $\wt{\Th_\phi\Th_\psi}$, which of course reduces to
(\ref{tag:ASY}) upon taking for $\psi$ the constant function equal to~$I$.
The~strange-looking operators $\MM _{mq}$ originate from the
formula~(\ref{tag:QL}).

\begin{theorem}\label{thm:TPT} For any functions $\phi,\psi\in C^\infty_
\CCNN(\OmegaN)$ and a diagonal matrix $X=\diag(c_1,\dots,c_N)$,
$$ [\wt{\Th_\phi \Th_\psi}(X)]_{ab} = O(h^\infty) \qquad\text{as } h\to0  $$
if $c_a\neq c_b$;
\begin{equation}  \begin{aligned} \relax
[\wt{\Th_\phi \Th_\psi}(X)]_{ab}
&\approx \sum \Sb i,j,k,l,\\ m,p,q=1\endSb ^N \sumr \; \frac{h^r}{r!} \;
\int_\UN \int_\UN u_{am} \ou_{im} u_{jm} \ou_{pm} w_{pq} \ow_{kq} \\
& \qquad\quad \cdot\, w_{lq} \ow_{bq} \; (\MM_{mq}^r \phi_{ij}
\psi_{kl}) \vphantom{\int_\UN}
(U,W; c_a\chi_m,c_a\chi_q) \, dU \, dW   \end{aligned}   \label{tag:TPA}
\end{equation}
as~$h\to0$ if $c_a=c_b\neq0$, where
$$ (\MM_{mq}^r \phi_{ij} \psi_{kl})(U,W;d,e) := \bigg[ \bigg( \Delta_{(d)} +
\Delta_{(e)} + \dfrac{\partial^2}{\partial d_m \partial \overline e_q} \bigg)
^r  \phi_{ij}(U;d) \; \psi_{kl}(W; e) \bigg] ;  $$
and
\begin{equation}  \begin{aligned} \relax
[\wt{\Th_\phi \Th_\psi}(X)]_{ab}
&\approx \sum\Sb i,j,k,l,m,\\p,q,L,M=1\endSb ^N \sumr
\; \frac{h^r}{r!} \; \int_\UN \int_\UN u_{aL} \ou_{iL} u_{jm} \ou_{pm}  \\
& \qquad\quad\vphantom{\int_\UN} \cdot\,
w_{pq} \ow_{kq} w_{lM} \ow_{bM} \; (\MM_{mq}^r \phi_{ij} \psi_{kl})(U,W;0,0)
\,dU \,dW   \end{aligned}   \label{tag:TPB}  \end{equation}
as $h\to0$ if $c_a=c_b=0$.   \end{theorem}

The formula (\ref{tag:TPB}) can clearly be simplified upon carrying out the
summations over $L$ and~$M$ and performing the two integrations (which can be
done since $\phi(U;0)=\phi(0)$ and $\psi(W;0)=\psi(0)$ are independent of $U$
and~$W$) via Schur's orthogonality relations; the~result~is
$$ [\wt{\Th_\phi \Th_\psi}(X)]_{ab} \approx
\sum_{m,q=1}^N \sumr \; \frac{h^r}{r!} \;
\MM_{mq}^r (\phi\psi)_{ab}(0,0).  $$
Similarly, as~with (\ref{tag:ASY}) and~(\ref{tag:TSS}), using (\ref{KAPP}) it
can be shown that for $c_a=0$ the formula (\ref{tag:TPA}) reduces~to
$$ [\wt{\Th_\phi \Th_\psi}(X)]_{ab} \approx \sum_{m,q=1}^N \sumr \;
\frac{h^r}{r!} \; \frac{\MM_{mq}^r [(\phi+I\Tr\phi)(\psi+I\Tr\psi)]_{ab}(0,0)}
{(N+1)^2},  $$
which is different from~(\ref{tag:TPB}). We~refrain from going into these
details because they are not needed anywhere in the sequel.

We~conclude this section by observing that (\ref{tag:TF}) also fails in
general.

\begin{proposition}\label{thm:NULO} Let $\phi$ be the $\CCNN$-valued function
on $\OmegaN$, $N\ge2$, defined~by $\phi(Z)=|\det Z|^2 I$. Then
$$ \lim_{h\to0} \wt{\Th_\phi}(X) =0 \qquad\forall X\in\OmegaN.  $$
\end{proposition}

\begin{proof} Since $\phi(VZV^*)=V\phi(Z)V^*$ $\forall V\in\UN$, it~is enough
to check the assertion for diagonal~$X$, so~let $X=\diag(c_1,\dots,c_N)$.
As~$\phi_{jk}(U;d)=\delta_{jk} |d_1\dots d_N|^2$, we~have
\begin{align*}
(\Delta_{(d)} \phi_{jk})(U;d)
&= \delta_{jk} \, \sum_m |d_1\dots \hat d_m \dots d_N|^2 ,   \\
(\Delta_{(d)}^2 \phi_{jk})(U;d)
&= \delta_{jk} \, \sum_{m\neq n} |d_1\dots \hat d_m \dots
\hat d_n \dots d_N|^2   \\
&= \delta_{jk} \, 2\sum_{m_1<m_2} |d_1\dots \hat d_{m_1} \dots
\hat d_{m_2} \dots d_N|^2 ,   \\
& \vdots  \\
(\Delta_{(d)}^r \phi_{jk})(U;d)
&= \delta_{jk} \, r! \, \sum_{m_1<m_2<\dots<m_r} |d_1\dots \hat d_{m_1} \dots
\hat d_{m_2} \dots \hat d_{m_r} \dots d_N|^2 ,   \\
(\Delta_{(d)}^r \phi_{jk})(U;d) &= 0 \qquad\text{for } r>N.
\vphantom{\sum^m}   \end{align*}
(Here the hat $\,\widehat{\;}\;$ indicates that the corresponding variable is
omitted.) Thus by~(\ref{tag:TDE}) and~(\ref{tag:TSS})
$$ [\wt{\Th_\phi}]_{ab} = \delta_{ab} (|c_a|^2 h^{N-1} + h^N) + O(h^\infty), $$
that is,
$$ \wt{\Th_\phi}(X) = h^{N-1} X^*X + h^N + O(h^\infty)  $$
as $h\to0$, and the assertion follows.      \end{proof}

Similarly, it~can be shown that (\ref{tag:TNUL}) breaks down too: for instance,
for $\phi(Z) = |\det Z|^2 e^{-\Tr(Z^*Z)}I$, one~has
$$ \|\Th_\phi\| \approx \|\phi\|_\infty^{1/N} h^{N-1} \quad\text{as }h\to0  $$
(where $\|\phi\|_\infty:=\sup_{Z\in\OmegaN} \|\phi(Z)\|_{\CCN\to\CCN}$). This
can be proved by observing that the operator $T_\phi$ is diagonal with respect
to the basis~(\ref{tag:ZK}), with eigenvalues
$$ \frac{(k+1)h^N}{(h+1)^{2N+k}};  $$
and $\sup_k (k+1)/(h+1)^k \approx 1/(eh) = \|\phi\|_\infty^{1/N}/h$. We~omit
the details.

We~now turn to classes of observables $\phi$ which are more manageable than the
general case.

\section{Spectral and $U$-invariant functions}\label{SUIF}
A~function $\phi(Z)$ of $Z\in\OmegaN$ will be called \emph{spectral} if it is a
function of $Z$ in the sense of the Spectral Theorem: that~is, if~there exists
a function $f:\CC\to\CC$ such that $\phi=f^\#$, where
\begin{equation}  f^\#(Z) := U \cdot \diag_j(f(d_j)) \cdot U^*
\qquad\text{if}\quad Z=U\cdot\diag_j(d_j)\cdot U^*.  \label{tag:GH}
\end{equation}
Our first observation is that for spectral functions, all goes fine with the
Berezin-Toeplitz quantization.

\begin{theorem}\label{thm:THX} If~$\phi=f^\#$ and $\psi=g^\#$ are two smooth
spectral functions, then there exist unique spectral functions $\rho_r$,
$r=0,1,2,\dots$, such that
$$ \Th_\phi\Th_\psi \approx \sumr \,h^r\;\Th_{\rho_r} \qquad\text{as }h\to0  $$
in~the sense of operator norms $($i.e.~as~in~$(\ref{tag:TC}))$. In~fact,
$$ \rho_r = C_r(f,g)^\# ,  $$
where
\begin{equation} C_r(f,g)=\frac1{r!}\,\partial^r f\cdot\dbar^r g
\label{tag:CR}  \end{equation}
are the operators $(\ref{tag:CF})$ for $n=1$.  \end{theorem}

\begin{proof} Recall that the monomials $z^k$, $k=0,1,2,\dots$, are orthogonal
in the Segal-Bargmann space~(\ref{tag:SB}) for $n=1$:
$$ \spr{z^k,z^l} _{L^2_\hol(\CC,d\mu_h)} = \delta_{kl} k! h^k.  $$
Comparing this with~(\ref{tag:UOG}), we~see that the mapping
\begin{equation} \iota: Z^k\chi_j \longmapsto z^k \otimes \chi_j
\label{isom}
\end{equation}
is a unitary isomorphism of our space $\HH_h$ onto the tensor product
$L^2_\hol(\CC,d\mu_h) \otimes \CCN$. Now if $\phi=f^\#$ is a spectral function
and $\chi,\eta\in\CCN$, then
\begin{align*}
\spr{ \Th_\phi Z^k \chi, Z^l\eta}
&= \spr{\phi Z^k\chi,Z^l\eta}_{L^2_\CCN(\OmegaN,d\mu_h)}  \\
&= \intoN \eta^* Z^{*l} \phi(Z) Z^k \chi \,d\mu_h(Z)  \\
&= \int_\CCN \int_\UN \eta^* U D^{*l} \phi(D) D^k U^*\chi \,dU \,
 e^{-\Tr(D^*D)/h} \, \frac{dD}{(\pi h)^N} .   \end{align*}
However, for any matrix~$X$,
\begin{equation}  \int_\UN UXU^* \, dU = \frac{\Tr(X)}N \, I.  \label{tag:UTR}
\end{equation}
(Indeed, performing the change of variable $U\mapsto U_1U$ and using the
invariance of the Haar measure, it~transpires that the left-hand side commutes
with any $U_1\in\UN$. Thus it must be a multiple of the identity. Taking traces
and using the cyclicity of the trace, (\ref{tag:UTR})~follows.) Thus we
can continue the above calculation with
\begin{align*}
\hskip4em
&= \spr{\chi,\eta} \;\frac1N
\int_\CCN \Tr(D^{*l} \phi(D) D^k) e^{-\Tr(D^*D)/h} \, \frac{dD}{(\pi h)^N} \\
&= \spr{\chi,\eta} \; \frac1N \sum_{j=1}^N \int_\CCN
\od_j^l d_j^k f(d_j) \, e^{-\|d\|^2/h} \, \frac{dD}{(\pi h)^N}  \\
&= \spr{\chi,\eta} \; \frac1N \sum_{j=1}^N \spr{z^k f,z^l}_{L^2(\CC,d\mu_h)} \\
&= \spr{\chi,\eta} \; \spr{\Th_f z^k,z^l}_{L^2_\hol(\CC,d\mu_h)}  \\
&= \spr{ (\Th_f \otimes I)(z^k\otimes\chi),z^l\otimes\eta}
_{L^2_\hol(\CC,d\mu_h)\otimes\CCN}.   \end{align*}
Consequently, under the isomorphism~$\iota$, the operator $\Th_\phi$ on $\HH_h$
corresponds to the operator $\Th_f\otimes I$ on $L^2_\hol(\CC,d\mu_h)\otimes
\CCN$, and the desired assertions follow immediately from the ordinary
Berezin-Toeplitz quantization on~$\CC$.  \end{proof}

We~list one more corollary of the above isomorphism~$\iota$; it~will not be
needed in the sequel, but should be contrasted with
Proposition~{\ref{thm:NULO}} at the end of Section~\ref{STPH} and the example
immediately thereafter. We~omit the proof.

\begin{proposition}\label{thm:NULP} For any spectral function $\phi=f^\#$ and
$x\in\CC$,
\begin{equation}  \wt{\Th_\phi} (xI) = \wt{\Th_f}(x) \cdot I   \label{tag:SBER}
\end{equation}
where the $\wt{\Th_f}$ on the right-hand side is the ordinary scalar-valued
Berezin transform of the operator $\Th_f$ on $L^2_\hol(\CC,d\mu_h)$.
In~particular,
$$ \lim_{h\to0} \|\wt{\Th_\phi}\|_\infty = \lim_{h\to0} \|\Th_\phi\|
=\|\phi\|_\infty.  $$   \end{proposition}

\begin{remark*} We~pause to note that for the full matrix domain
$\Omega=\CCNN$, the spaces $\HH_h$ are not isomorphic to $L^2_\hol(\CC,d\nu)$
for any rotation invariant measure~$\nu$ on~$\CC$. The reason is that the
numbers $c_k$ in~(\ref{tag:CK}), which take over the role of the~$k!$, are not
the moment sequence of any measure on $[0,\infty)$ if $N>1$. This can be seen
by checking that
$$ \frac1{(k+1)(k+2)}\prod_{j=1}^{k+1}(N+j) = \int_\CC |z|^{2k} \,d\nu_N(z)  $$
where
$$ d\nu_N(z) := \frac1\pi \sum_{j=0}^{N-1} \frac{(N-1)!(N-j)}{j!}
\; |z|^{2j}\,e^{-|z|^2} \,dz;  $$
thus for $c_k$ to be a moment sequence (even of a measure which is not
necessarily non-negative) it is necessary and sufficient that
\begin{equation}  \bigg\{ \frac1{(k+1)(k+2)}\,\prod_{j=1}^{k+1} (N-j)
\bigg\}_{k=1}^\infty  \label{tag:TRH}  \end{equation}
be a moment sequence. However, the latter cannot be the case, since
(\ref{tag:TRH}) has only a finite number of nonzero terms.

We~restrict our attention exclusively to $\OmegaN$ in the rest of this paper.
\qed   \end{remark*}

\medskip

Returning to the main line of discussion, we~proceed to introduce another class
of functions.

A~$\CCNN$-valued function $\phi$ on $\OmegaN$ will be called
\emph{$U$-invariant}~if
\begin{equation}  \phi(UZU^*) = U\,\phi(Z) \,U^* \qquad\forall U\in\UN
 \;\forall Z\in\OmegaN.  \label{tag:UI}   \end{equation}
Clearly, a~spectral function is $U$-invariant, but not vice versa: an~example
is the function $\phi(Z)=|\det Z|^2 I$ from the end of Section~\ref{STPH}. The
relationship between spectral and $U$-invariant functions is clarified in the
next proposition.

\begin{proposition}\label{thm:SPE} A~function $\phi$ is $U$-invariant
if and only if there exists a function $f(d_1;d_2,\dots,d_N)$ from
$\CC\times\CC^{N-1}$ into~$\CC$, symmetric in the $N-1$ variables $d_2,\dots,
d_N$, such that $\phi=f^\#$, where
\begin{equation}  f^\# (UDU^*) := U\cdot \diag_j(f(d_j;d_1,\dots,\hat
d_j,\dots,d_N))  \cdot U^*.  \label{tag:UJ}  \end{equation}
The function $f$ is uniquely determined by~$\phi$.

Further, $\phi$ is spectral if and only if $f$ depends only on the first
variable, i.e.~if and only if $f(d_1;d_2,\dots,d_N)=f(d_1;0,\dots,0)$.
\end{proposition}

\begin{proof} For any complex numbers $\epsilon_1,\dots,\epsilon_N$ of
modulus~one, consider the matrix $\bep=\diag(\epsilon_1,\dots,\epsilon_N)$.
Then $\bep\in\UN$ and $\bep D\bep^*=D$ for any diagonal matrix~$D$; thus
by~(\ref{tag:UI})
$$ \phi(D) = \bep \,\phi(D) \, \bep^* \qquad \forall
\epsilon_1,\dots,\epsilon_N\in\mathbf T.  $$
Consequently, $\phi(D)$ is also a diagonal matrix. Define the functions
$f_1,\dots,f_N$ on $\CCN$~by
\begin{equation}  f_j(d_1;d_2,\dots,d_N) := \phi_{jj}(D) \qquad\text{where }
D=\diag(d_1,\dots,d_N).  \label{tag:UK}  \end{equation}
For any permutation $\sigma$ of the set $\{1,2,\dots,N\}$, let $F_\sigma$
denote the permutation matrix $[F_\sigma]_{jk}=\delta_{\sigma(j),k}$. Then
$F_\sigma\in\UN$ and
$$ F_\sigma D F_\sigma^* = \diag(d_{\sigma(1)},\dots,d_{\sigma(N)}) \qquad
\text{if } D=\diag(d_1,\dots,d_N).  $$
Thus by (\ref{tag:UI}) again
$$ f_{\sigma(j)}(d_1;d_2,\dots,d_N) = f_j(d_{\sigma(1)};d_{\sigma(2)},\dots,
d_{\sigma(N)}).  $$
It~follows that $f_j$ is symmetric with respect to the $N-1$ variables
$d_1,\dots,\hat d_j,\dots,d_N$ and $\phi=f^\#$ for $f=f_1$.

Conversely, it~is easily seen that any function of the form (\ref{tag:UJ}) is
$U$-invariant, and $f^\#=g^\#\iff f=g$.

Finally, the assertion concerning spectral functions is immediate upon
comparing (\ref{tag:UJ}) and~(\ref{tag:GH}).      \end{proof}

One consequence of the last proposition is that the mapping
\begin{equation}  f^\# \longmapsto (f^\flat)^\#   \label{tag:KA}
\end{equation}
with $f^\flat:\CC\to\CC$ defined~by
$$ f^\flat(z) := f(z;0,\dots,0)  $$
is~a projection from $U$-invariant functions onto spectral functions.
(Here the first $\#$ in (\ref{tag:KA}) is the one for $U$-invariant functions
from~(\ref{tag:UJ}), while the second is the one for spectral functions
from~(\ref{tag:GH}); however, there is no danger of confusion in this abuse
of notation.)  In~terms of~$f^\#=\phi$, the function $f^\flat$ can be expressed
directly~by
$$ f^\flat(z) = \phi_{11}(zE_{11}), \qquad z\in\CC,  $$
where $E_{11}$ is the matrix of projection onto the first coordinate,
i.e.~$[E_{11}]_{jk} = \delta_{1j}\delta_{1k}$. The~projections $f\mapsto
f^\flat$ and (\ref{tag:KA}) will play a crucial role in the next section.

\section{Quantization of $U$-invariant functions}\label{QUIF}
We~now proceed to establish our final result --- a~generalization of
Theorem~{\ref{thm:THX}} to $U$-invariant functions. The key ingredient is
played by the following specializations of the asymptotic expansions from
Section~\ref{STPH}.

\begin{theorem}\label{thm:ASS} For any smooth $U$-invariant functions
$\phi=f^\#$ and $\psi=g^\#$ on~$\OmegaN$,
\begin{equation}  \wt{\Th_\phi} \approx \sumr \, h^r \, (l_r\phi)^\#
\label{tag:KC}  \end{equation}
and
\begin{equation}  \wt{\Th_\phi\Th_\psi} \approx \sumr \, h^r \;
m_r(\phi,\psi)^\#  \label{tag:KD}  \end{equation}
as $h\to0$, where $l_r\phi$ and $m_r(\phi,\psi)$ are the functions on $\CC$
defined~by
\begin{equation}  l_r \phi(z) := \frac1{r!} \, (\Delta^r f)(z;0,\dots,0)
\label{tag:KE}  \end{equation}
$($that~is, $l_r\phi=\frac1{r!}(\Delta^r_{(d)}\phi)^\flat)$ and
\begin{equation}  m_r(\phi,\psi)(z) := \bigg[ \frac1{r!} \, \bigg( \Delta_{(d)}
+ \Delta_{(e)} + \frac{\partial^2}{\partial d_1\partial\overline e_1} \bigg)^r
f(d) g(e) \bigg] \bigg| \Sb d=(z;0,\dots,0) \\ e=(z;0,\dots,0) \endSb.
\label{tag:KF}  \end{equation}  \end{theorem}

\begin{proof} In~principle this could be gleaned from the formulas
(\ref{tag:ASY}) and~(\ref{tag:TSS}), but it is better to use directly the
definitions: if~$X=VCV^*$ with $V\in\UN$ and $C=\diag(c_1,\dots,c_N)$, then by
Proposition~{\ref{thm:PRB}}
\begin{align*}
&\wt{\Th_\phi}(X) = \intoN K_h(X,X)^{-1/2} K_h(X,Y) \phi(Y) K_h(Y,X)
K_h(X,X)^{-1/2} \,d\mu_h(Y)  \\
&\quad= \int_\CCN \int_\UN  V K_h(C,C)^{-1/2} V^* \sum_{k=0}^\infty
\frac{VC^kV^*UD^{*k}U^*}{k!h^k} \, \phi(UDU^*)   \\
&\hskip8em   \cdot\,\sum_{l=0}^\infty
\frac{UD^lU^*VC^{*l}V^*}{l!h^l} \, V K_h(C,C) V^* \,dU
\, e^{-\Tr(D^*D)/h} \, \frac{dD}{(\pi h)^N}  \\
&\quad= \int_\CCN \int_\UN  V e^{-CC^*/2h}  \sum_k \frac{C^kV^*UD^{*k}}{k!h^k}
\, \phi(D)   \\
&\hskip8em   \cdot\,\sum_l \frac{D^lU^*VC^{*l}}{l!h^l} \, e^{-CC^*/2h} V^*
\,dU \, e^{-\Tr(D^*D)/h} \, \frac{dD}{(\pi h)^N}  \\
&\quad= \frac1N \int_\CCN V e^{-CC^*/2h} \sum_{k,l}
\frac{C^kC^{*l}}{k!l!h^{k+l}} \; \Tr(D^{*k} \phi(D) D^l) \\
&\hskip8em    \cdot\, e^{-CC^*/2h} V^* \; e^{-\Tr(D^*D)/h}
\, \frac{dD}{(\pi h)^N}  \\
&\quad= \frac1N \sum_{j=1}^N \int_\CCN V e^{-CC^*/h} e^{(\od_j C+d_j C^*)/h}
\phi_{jj}(D) \, V^* \, e^{-\|d\|^2/h}  \, \frac{dD}{(\pi h)^N}   \\
&\quad= V \cdot \diag_k \bigg( \frac1N \sum_{j=1}^N \int_\CCN
f(d_j;d_1,\dots,\hat d_j,\dots,d_N) \, e^{-\|c_k \chi_j-d\|^2/h}
\, \frac{dD}{(\pi h)^N} \bigg) \cdot V^*  \\
&\quad= V \cdot \diag_k \bigg( \int_\CCN f(d_1;d_2,\dots,d_N) \,
e^{-\|c_k\chi_1-d\|^2/h} \, \frac{dD}{(\pi h)^N} \bigg) \cdot V^*  \\
&\quad= V \cdot \diag_k \bigg( \sumr \, \frac{h^r}{r!} \,
(\Delta^r f)(c_k;0,\dots,0) \bigg) \cdot V^*  \\
&\quad= \sumr \, h^r \; (l_r\phi)^\# (X).   \end{align*}
Here we have used, in~turn, the formula~(\ref{tag:KN}) for $K_h(X,Y)$; the
$U$-invariance of~$\phi$; the formula (\ref{tag:UTR}) for the integral
over~$\UN$; the fact that $\Tr(D^{*k}\phi(D) D^l)=\sum_{j=1}^N \od_j^k d_j^l
\phi_{jj}(D)$, combined with the summation of the exponential series and the
commutativity of $C$ with~$C^*$; the fact that $\phi=f^\#$; the independence of
the integral on~$j$; the stationary phase expansion; and (\ref{tag:GH}) and the
definition of~$l_r\phi$.

The proof of (\ref{tag:KD}) is similar:\footnote{Of~course, (\ref{tag:KC})~can
also be obtained from (\ref{tag:KD}) upon setting $\psi\equiv I$; but it is
more instructive to give a separate proof.}
\begin{align*}
& \wt{\Th_\phi\Th_\psi}(X) = \intoN \intoN K_h(X,X)^{-1/2} K_h(X,Y) \phi(Y)
K_h(Y,Z) \psi(Z) \\
&\hskip8em \vphantom{\int_\CCN}  \cdot\,
K_h(Z,X) K_h(X,X)^{-1/2} \,d\mu_h(Y) \,d\mu_h(Z)  \\
&\quad= \int_\CCN \int_\CCN \int_\UN \int_\UN V e^{-CC^*/2h} \sum_k
\frac{C^kV^*UD^{*k}}{k!h^k}  \, \phi(D) \sum_l \frac{D^lU^*WE^{*l}}{l!h^l} \\
&\hskip8em  \cdot\, \psi(E) \sum_m \frac{E^mW^*VC^{*m}}{m!h^m} \,e^{-CC^*/2h}
V^* \,dU\,dW \,d\mu_h(D) \,d\mu_h(E)  \\
&\quad= \frac1{N^2} \int_\CCN \int_\CCN V e^{-CC^*/2h} \sum_{k,l,m}
\frac{C^kC^{*m}}{l!k!m!h^{k+l+m}} \,\Tr(D^{*k}\phi(D)D^l)   \\
&\hskip8em \vphantom{\int_\CCN} \cdot\,
\Tr(E^{*l}\psi(E)E^m) \, d\mu_h(D) \, d\mu_h(E)  \\
&\quad= \frac1{N^2} \int_\CCN \int_\CCN  \sum_{i,j=1}^N V e^{-CC^*/h}
e^{(\od_iC+d_i\overline e_j+e_jC^*)/h} \phi_{ii}(D) \psi_{jj}(E)
\,d\mu_h(D) \,d\mu_h(E)  \\
&\quad= V \cdot \diag_k \bigg( \frac1{N^2} \sum_{i,j} \int_\CCN \int_\CCN
e^{-|c_k|^2/h} e^{(\od_ic_k+d_i\overline e_j+e_j\oc_k)/h} \\
&\hskip8em  \cdot\,
e^{-(\|d\|^2+\|e\|^2)/h} \phi_{ii}(D)\psi_{jj}(E)
\, \frac{dD}{(\pi h)^N} \, \frac{dE}{(\pi h)^N} \bigg) \cdot V^* \\
&\quad= V \cdot \diag_k \bigg( \int_\CCN \int_\CCN
e^{-|c_k|^2/h} e^{(\od_1 c_k+d_1\overline e_1+e_1\oc_k)/h}
e^{-(\|d\|^2+\|e\|^2)/h}  \\
&\hskip8em  \cdot\,  f(d_1;d_2,\dots,d_N) g(e_1;e_2,\dots,e_N)
\, \frac{dD}{(\pi h)^N} \,\frac{dE}{(\pi h)^N} \bigg) \cdot V^*  \\
&\quad= V \cdot \diag_k \bigg( \sumr \, \frac{h^r}{r!} \; \Big[ \Delta_{(d)}+
\Delta_{(e)} + \frac{\partial^2}{\partial d_1\partial\overline e_1} \Big]^r
f(d) g(e) \Big|\Sb d=(c_k;0,\dots,0)\\e=(c_k;0,\dots,0)\endSb \bigg)
\cdot V^*  \\
&\quad= \sumr \, h^r \; (m_r(\phi,\psi))^\# (X).   \end{align*}
Here the penultimate line comes from the formula~(\ref{tag:QL}).
\end{proof}

\begin{corollary}\label{thm:COF} If~$g$ is a smooth function on $\CC$ and
$g^\#$ the corresponding spectral function on~$\OmegaN$, then
\begin{equation}  \wt{\Th_{g^\#}} \approx \sumr \, \frac{h^r}{r!} \; (\Delta^r
g)^\# \qquad\text{as } h\to0.  \label{tag:SPA}  \end{equation}
In~particular,
\begin{equation}  \lim_{h\to0} \wt{\Th_{g^\#}}(X)=0\ \forall X \iff g\equiv0.
\label{tag:UNQ}  \end{equation}  \end{corollary}

\begin{proof} Combine (\ref{tag:KE}) with the last part of
Proposition~{\ref{thm:SPE}}.   \end{proof}

The~following theorem is the main result of this section and indeed, of this
paper.

\begin{theorem}\label{thm:MAT} For any smooth $U$-invariant functions
$\phi,\psi$ on~$\OmegaN$, there exist uniquely determined functions
$g_0,g_1,\dots$, on~$\CC$ such that
\begin{equation} \wt{\Th_\phi \Th_\psi} \approx \sum_{m=0}^\infty h^m
\;\wt{\Th_{g^\#_m}} \qquad\text{as }h\to0.  \label{ASEX}  \end{equation}
Moreover, if $\phi=f^\#$ and $\psi=g^\#$, then the functions $g_m$ are given~by
\begin{equation}  g_m = G_m(f,g)^\flat  \label{tag:GO}  \end{equation}
for some bidifferential operators $G_m$ on~$\CCN$ $($independent of~$f$
and~$g)$. In~particular,
\begin{equation}
\begin{gathered}  G_0(f,g)^\flat = f^\flat g^\flat,  \qquad\text{and}  \\
G_1(f,g)^\flat - G_1(g,f)^\flat = \frac i{2\pi} \{f^\flat,g^\flat\},
\end{gathered}  \label{tag:GP}   \end{equation}
the Poisson bracket of $f^\flat$ and $g^\flat$ on~$\CC$.   \end{theorem}

\begin{proof} The uniqueness is immediate from~(\ref{tag:UNQ}). The
existence~is, by~virtue of~(\ref{tag:KD}) and~(\ref{tag:SPA}), equivalent~to
$$ \sumr h^r \, m_r(\phi,\psi)^\# \approx \sum_{m,n=0}^\infty h^{m+n}
\frac{(\Delta^n g_m)^\#}{n!}.  $$
Comparing the expressions at like powers of $h$ on both sides, this becomes
$$ m_r(\phi,\psi) = \sum_{n=0}^r \frac{\Delta^n g_{r-n}}{n!},  $$
which is solved by the recursive recipe
\begin{equation}  g_r = m_r(\phi,\psi) - \sum_{n=1}^r\frac1{n!}\,\Delta^n
g_{r-n}. \label{tag:GG}  \end{equation}
From (\ref{tag:KF}) it is also clear that $g_m$ are of the form~(\ref{tag:GO})
with appropriate bidifferential operators~$G_m$. Finally, a~short computation
using the special instances $r=0,1$ of~(\ref{tag:KF}),
$$ m_0(\phi,\psi)=f^\flat g^\flat, \qquad
 m_1(\phi,\psi)=\Big( g\Delta f + f\Delta g + \frac{\partial f}{\partial d_1}
\frac{\partial g}{\partial\overline e_1} \Big)^\flat ,   $$
gives~(\ref{tag:GP}).     \end{proof}

\begin{remark*} Note that the quantities $G_m(\phi,\psi)^\flat$ do \emph{not}
depend only on $f^\flat$ and~$g^\flat$: the~bidifferential operators $G_m$
involve derivatives also in other variables than $d_1,e_1$, and only after
these are applied one takes the restriction to $d_2=\dots=d_N=e_2=\dots=e_N=0$.
It~is therefore quite remarkable that $G_1(\phi,\psi)^\flat-G_1(\psi,\phi)
^\flat $ depends only on $f^\flat$ and $g^\flat$ --- the derivatives with
respect to the other variables having cancelled~out.  \qed   \end{remark*}

We~indicate another proof of the last theorem, based on the
isomorphism~(\ref{isom}). (We~gave the proof above first since the isomorphism
(\ref{isom}) is probably something peculiar to the domain of normal matrices,
while the stationary phase method should work also in other situations.
The~proof below also requires a slightly stronger hypothesis on the
functions~$\phi$ and~$\psi$.)

For~a~function $f$ on $\CCN$ and $h>0$, let $P_h f$ be the function on~$\CC$
defined~by
$$ P_h f(z_1) := \int_{\CC^{N-1}} f(z_1,z_2,\dots,z_N) \;
e^{-(|z_2|^2+\dots+|z_N|^2)/h} \; \frac{dz_2\dots dz_N}{(\pi h)^{N-1}}.  $$

\begin{theorem} \label{projthm} Let $\phi=f^\#$, $\psi=g^\#$ be smooth
$U$-invariant functions on~$\OmegaN$ such that the partial derivatives of
$f$ and $g$ of all orders are bounded, and let $C_r$ be the bidifferential
operators~$(\ref{tag:CR})$. Then
$$ \Th_\phi\Th_\psi \approx \sumr\; h^r \, \Th_{C_r(P_h f,P_h g)^\#}  $$
in the sense of operator norms. Consequently, $(\ref{ASEX})$~holds for
$$ g_m = \sum\Sb j,k,r\ge0,\\ j+k+r=m\endSb \frac1{j!k!r!} \, \partial^r
(\Delta^{\prime j}f)^\flat \cdot \dbar^r(\Delta^{\prime k}g)^\flat ,  $$
where $\Delta'$ denotes the Laplacian with respect to the last $N-1$ variables
$z_2,\dots,z_N$.   \end{theorem}

\begin{proof} By~a~computation similar to the one in the proof of
Theorem~\ref{thm:THX}, for any $\chi,\eta\in\CCN$,
\begin{align*}
\spr{ \Th_\phi Z^k \chi, Z^l\eta}
&= \intoN \eta^* Z^{*l} \phi(Z) Z^k \chi \,d\mu_h(Z)  \\
&= \int_\CCN \int_\UN \eta^* U D^{*l} U^* \phi(UDU^*) U D^k U^*\chi \,dU \,
 e^{-\Tr(D^*D)/h} \, \frac{dD}{(\pi h)^N} \\
&= \int_\CCN \int_\UN \eta^* U D^{*l} \phi(D) D^k U^*\chi \,dU \,
 e^{-\Tr(D^*D)/h} \, \frac{dD}{(\pi h)^N}  \\
& \hskip18em \text{(by the $U$-invariance of $\phi$)}  \\
&= \spr{\chi,\eta} \;\frac1N
\int_\CCN \Tr(D^{*l} \phi(D) D^k) e^{-\Tr(D^*D)/h} \, \frac{dD}{(\pi h)^N} \\
&= \spr{\chi,\eta} \; \frac1N \sum_{j=1}^N \int_\CCN
\od_j^l d_j^k f(d_j;d_1,\dots,\hat d_j,\dots,d_N) \, e^{-\|d\|^2/h} \,
\frac{dD}{(\pi h)^N}  \\
&= \spr{\chi,\eta} \; \int_\CCN
\od_1^l d_1^k f(d_1;d_2,\dots,\dots,d_N) \, e^{-\|d\|^2/h} \,
\frac{dD}{(\pi h)^N}   \\
&= \spr{\chi,\eta} \; \int_\CC
\od_1^l d_1^k P_h f(d_1) \, e^{-|d_1|^2/h} \, \frac{d d_1}{\pi h} \\
&= \spr{\chi,\eta} \; \spr{z^k P_h f,z^l}_{L^2(\CC,d\mu_h)} \\
&= \spr{\chi,\eta} \; \spr{\Th_{P_h f} z^k,z^l}_{L^2_\hol(\CC,d\mu_h)}  \\
&= \spr{ (\Th_{P_h f} \otimes I)(z^k\otimes\chi),z^l\otimes\eta}
_{L^2_\hol(\CC,d\mu_h)\otimes\CCN}.   \end{align*}
Consequently, under the isomorphism~(\ref{isom}), the operator $\Th_\phi$ on
$\HH_h$ corresponds to the operator $\Th_{P_h f}\otimes I$ on $L^2_\hol(\CC,
d\mu_h)\otimes\CCN$. Thus by the ordinary Berezin-Toeplitz quantization
on~$\CC$,
\begin{align*}
\Th_\phi\Th_\psi &\cong \Th_{P_h f}\Th_{P_h g} \otimes I \\
&\approx \sumr h^r \, \Th_{C_r(P_h f,P_h g)} \otimes I \\
&\cong \sumr h^r \, \Th_{C_r(P_h f,P_h g)^\#}  \end{align*}
(the~last isomorphism is the one from the proof of Theorem~\ref{thm:THX}).
This proves the first claim. The second part of the theorem follows upon
inserting the expansion
$$ P_h f =\sum_{j=0}^\infty \frac{h^j}{j!} \, (\Delta^{\prime j} f)^\flat,  $$
which follows from the Taylor formula (or~stationary phase), and taking Berezin
transforms on both sides. (The~hypothesis of boundedness of the derivatives
of $f$ and $g$ is needed in order that the resulting expansion for $C_r(P_h f,
P_h g)$ converge uniformly on~$\CCN$, and thus imply the convergence of the
corresponding expansion for $\Th_{C_r(P_h f,P_h g)^\#}$ by the inequality
$\|T_\phi\|\le\|\phi\|_\infty$.)   \end{proof}

\medskip

For~two $U$-invariant functions $\phi=f^\#$, $\psi=g^\#$, define their ``star
product'' $\phi*\psi$ as the formal power series
$$ \phi * \psi := \sumr h^r \, G_r(f,g)^{\flat\#}.  $$
As~usual, this product can be extended by $\CC[[h]]$-linearity to all $\phi,
\psi\in\mathcal U[[h]]$, the ring of all power series in $h$ with coefficients
in the algebra $\mathcal U$ of all $U$-invariant functions on~$\OmegaN$.
Alternatively, upon identifying $\phi=f^\#\in\mathcal U$ with~$f$, we~may view
this as the star product
$$ f * g := \sumr h^r \, G_r(f,g)^\flat  $$
on~the algebra $\mathcal S$ of all functions $f(d_1;d_2,\dots,d_N)$
on~$\CC\times\CC^{N-1}$ symmetric in the last $N-1$ variables, which again can
be extended by $\CC[[h]]$-linearity to all $f,g\in\mathcal S[[h]]$, the ring of
formal power series with coefficients in~$\mathcal S$. If~we extend to
$\mathcal S[[h]]$ by $\CC[[h]]$-linearity also the operators~$G_r$, then the
extended star-product will still satisfy the relations~(\ref{tag:GP}). Further,
$*$~is~clearly associative, since the multiplication of operators~is associative --- both
$(\phi*\psi)*\eta$ and $\phi*(\psi*\eta)$ originate from the asymptotic
expansion as $h\to0$ of $[\Th_\phi \Th_\psi \Th_\eta]\,\wt{\;}$.
(However, in~contrast to a genuine star-product, the function constant one is
\underbar{not} the unit element for~$*$.)

The~appearance of $f^\flat$ and~$g^\flat$, and not $f$ and~$g$,
in~(\ref{tag:GP}) means that the $\CC^{N-1}$ part of $f$ disappears in the
semiclassical limit $h\to0$, and only the projection~$f^\flat$, which lives
on~$\CC$, survives. As mentioned before, we are dealing here with a quantum
system which has $N$ internal degrees of freedom. This is made clear by the
isomorphism (\ref{isom}), since the tensor product space $L^2_\hol(\CC,d\mu_h)
\otimes \CCN$ is exactly the Hilbert space of a single quantum particle, moving
on the phase space $\CC$ and  having $N$ internal degrees of freedom. The full
set of quantum observables of this system include those which do not have
classical counterparts. The interesting fact that emerges from our analysis is
that, it is exactly those observables which are Berezin quantized versions of
$U$-invariant functions, that have classical counterparts. Since the internal
degrees of freedom are purely quantum in this case, they do not survive in the
semi-classical limit.

\bigskip

\noindent{\textsc{Acknowledgement.}} Part of this work was done while the
second author was visiting the first; the~support of the Mathematics Department
of Concordia University is gratefully acknowledged.

\end{document}